\newif\ifshort
\newif\ifveryshort
\newif\ifsup
\veryshortfalse

\shorttrue
\suptrue

\documentclass[journal]{IEEEtran}

\ifCLASSOPTIONcompsoc
  \usepackage[nocompress]{cite}
\else
  \usepackage{cite}
\fi

\usepackage{amsthm}
\usepackage{latexsym}
\usepackage{amssymb}
\usepackage{graphicx}
\usepackage{caption}
\usepackage{subcaption}
\usepackage{amsmath}
\usepackage{color}
\usepackage{cite}
\usepackage{cleveref}
\usepackage[english]{babel}

\theoremstyle{plain}

\newtheorem{lemma}{Lemma}

\newtheorem{corollary}{Corollary}
\theoremstyle{definition}
\newtheorem{definition}{Definition}
\theoremstyle{remark}

\newcommand{\off}[1]{}
\newcommand{\revised}[1]{\textcolor[rgb]{0.00,0.00,0.00}{#1}}

\newcommand{\revisedb}[1]{\textcolor[rgb]{0.00,0.00,0.00}{#1}}
\newcommand{\rev}[1]{\textcolor[rgb]{0.00,0.00,0.00}{#1}}
\newcommand{\offall}[1]{\off{}}

\ifCLASSINFOpdf
\else
\fi

\begin{document}
\title{Efficient Data Collection Over Multiple Access\\ Wireless Sensors Network
\thanks{A. Cohen, A. Cohen and O. Gurewitz are with the Department of Communication Systems Engineering, Ben-Gurion University of the Negev, Beer-Sheva 84105, Israel. E-mail: alejandr@post.bgu.ac.il; coasaf@bgu.ac.il; gurewitz@bgu.ac.il.
Parts of the secure protocol design were presented at the International Symposium on Cyber Security, Cryptography and Machine Learning, CSCML 2018. \rev{Patent application submitted No. 62/566,536.}}}

\author{\hspace{-1 mm}\IEEEauthorblockN{Alejandro Cohen,~\IEEEmembership{Student Member,~IEEE,} Asaf Cohen,~\IEEEmembership{Member,~IEEE,} and Omer Gurewitz,~\IEEEmembership{Member,~IEEE,\\\rev{In memory of Aviad Firuz (R.I.P.)}}}\hspace{-1 mm}\vspace{-1cm}}
\maketitle
\begin{abstract}
Data collection in Wireless Sensor Networks (WSN) draws significant attention, due to emerging interest in technologies ranging from Internet of Things (IoT) networks to simple ``Presence" applications, which identify the status of the devices (active or inactive). Numerous Medium Access Control (MAC) protocols for WSN, which can address the challenge of data collection in dense networks, were suggested over the years. Most of these protocols utilize the traditional layering approach, in which the MAC layer is unaware of the encapsulated packet payload, and therefore there is no connection between the data collected, the physical layer and the signaling mechanisms. Nonetheless, in many of the applications that intend to utilize such protocols, nodes may need to exchange very little information, and do so only sporadically, that is, while the number of devices in the network can be very large, only a subset wishes to transmit at any given time. Thus, a tailored protocol, which matches the signaling, physical layer and access control to traffic patterns is required.

In this work, we design and analyze a data collection protocol based on information theoretic principles. In the suggested protocol, the sink collects messages from up to $K$ sensors simultaneously, out of a large population of sensors, without knowing in advance which sensors will transmit, and without requiring any synchronization, coordination or management overhead. In other words, neither the sink nor the other sensors need to know who are the actively transmitting sensors, and this data is \textit{decoded} directly from the channel output. We provide a simple codebook construction with very simple encoding and decoding procedures. We further design a secure version of the protocol, in which an eavesdropper observing only partial information sent on the channel cannot gain significant information on the messages transmitted or even which are the sources that sent these messages.
\end{abstract}
          %
\vspace{-0.5cm}
\section{Introduction}\label{intro}
Utilizing scattered Wireless Sensor Networks (WSN) for gathering environmental information from a large number of sensor nodes with limited capabilities continues to draw a lot of attention from both industrial and academic communities, due to the large number of applications that rely on such infrastructures. For example, WSN will be a crucial technology enabler for implementation in the emerging Internet of Things (IoT) environment \cite{gubbi2013internet}, which will allow collection of information from densely deployed sensors, many of which are cheap with very limited capabilities (e.g., memory or energy resources) and compete for limited wireless resources (e.g., time or spectrum). In many of these applications, the majority of the sensors need to report only a limited amount of information, and do so only infrequently. For example, many of these sensors need to periodically send a keep alive message to inform the sink node that their battery has not drained out, and occasionally report one of several possible events, e.g., motion detected, temperature is above a given threshold, or one out of several quantized values. Accordingly, the main challenge for such networks is to cope with a huge number of simple devices that need to send limited information, competing for very limited wireless resources (compared to the number of sensors) while saving energy.

Numerous Medium Access Control (MAC) protocols for WSN have been suggested over the years, designed to cope with various setups and objectives (a short representative overview is provided in Section~\ref{related}). In particular, traffic in a WSN can be quite dynamic, depending on the events being sensed, the sensing application and the protocols being used. Therefore, such protocols should perform well under a wide range of traffic loads, a variety of network topologies and various objectives such as latency, reliability, energy consumption, security, etc. Accordingly, many of these MAC protocols were designed and examined to be robust under diverse setups. The approach of designing a MAC protocol regardless of the wireless Phy-protocol being used, the routing protocol employed, and even the application expected to utilize it, is consistent with the network-layering conceptual model. However, even though these protocols perform well under a large variety of setups, this universality comes at a price. For example, a MAC protocol allowing any subset of the sensors, regardless of its size, to transmit simultaneously will suffer from many collisions and high overhead if indeed a large subset attempts transmission concurrently. Another example is a \textit{keep alive} frame transmitted over WiFi, that conveys only the sender ID and a single bit of relevant information (the sensor is alive), yet requires a large number of bits to be transferred due to headers, physical encapsulation, etc.

In this paper, we design, analyze and evaluate a highly efficient WSN MAC protocol specially designed to collect information from a large number of sensors, utilizing information theoretic concepts and novel signaling and decoding techniques which allow us to jointly optimize all layers together. We assume the sensors are very simple, with highly constrained capabilities, e.g., no power control, rate adaptation mechanisms or sophisticated algorithmic capabilities. Thus, the key idea that our protocol relies on, is that instead of the typical frame mechanism using data encapsulation, each sensor is assigned a unique \textit{transmission pattern} for each of its messages, which conveys both the information and the sensor's ID. Whenever a sensor wishes to transmit a report, it waits to receive a predefined periodic preamble sent by its designated sink, and then transmits a sequence of \textit{impulses} according to the transmission pattern which corresponds to the report it wishes to send. The sink node \textit{can receive and decode several simultaneous transmissions in a way that it can recognize both the sender and the information sent} from the collected channel output received. \rev{Namely, a collision resolution procedure or scheduling (\hspace{-0.01mm}\cite{rom1990multiple}) are not required to decode the $K$ simultaneously transmission.} This is done using a carefully designed codebook, and a matching decoding algorithm that identifies both the sensors which transmitted as well as their codewords. Interestingly, unlike Code-Division Multiple-Access Channels (CDMA), the sink node can rely on a simple energy detection in order to decode, and not on the exact received power or any power adaptation mechanism, thus dramatically improving robustness.
\begin{figure}
  \centering
  \includegraphics[trim=0cm 0cm 0cm 0cm,clip,scale=0.3]{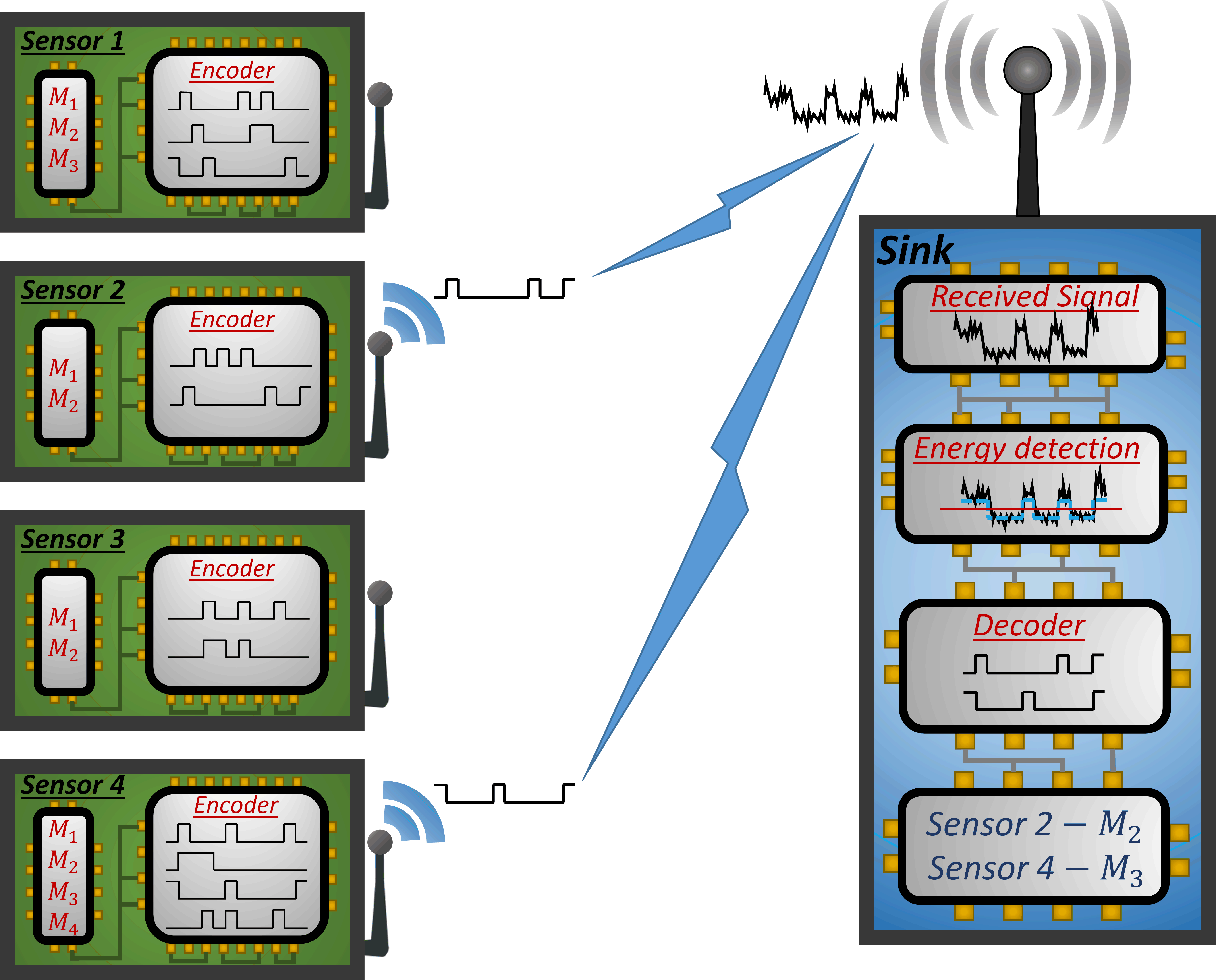}
  \caption{A schematic illustration of an IoT network. Data is collected by a sink, when $K$ sensors, out of a large population of $N$ sensors in the network have a message to transmit over the wireless channel.}
  \label{figure:WSN_model}
  \ifshort \vspace{-4.5mm} \fi
\end{figure}

To provide some insight into the suggested approach and illustrate its basic concept, consider the toy example depicted in Figure~\ref{figure:WSN_model}. In this example, we assume four sensors transmitting to a sink. The four sensors have $3, 2, 2$ and $4$ messages each. Each one of the messages is assigned a unique pattern, comprised of high and low level elements which are known to the sensor itself and to the sink. For example, Sensor $1$'s first message is encoded starting with a low level symbol followed by a high level symbol, a sequence of $4$ low level symbols, etc. After receiving a predefined beacon from the sink, which initiates a conceptual set of mini-timeslots, each sensor ready to transmit emits energy (transmits) according to the pattern assigned to the message to be transmitted. Specifically, it emits energy in the minislots which correspond to a high level in the message pattern and stays idle in the other minislots. In the above example only two sensors are awake and ready to transmit (sensors 2 and 4). Sensor 2 emits energy in minislots $2, 8$ and $11$ according to the pattern assigned to its second message, and Sensor 4 emits energy in minislots $1, 6$ and $11$ according to the pattern assigned to its third message. The received signal at the sink is a combination of the two transmitted sequences (energy at minislots $1, 2, 6, 8$ and $11$, with some additional noise). The sink performs an energy detection procedure on the received signal according to a predefined threshold, identifying which minislots were busy and which were idle (below the energy threshold). Note that due to the energy aggregate, the receiver gain can be saturated (e.g., several sensors emitted energy into the same minislot) yet all the sink needs to identify is on which minislots the energy is above the noise level (above the threshold). In other words, each minislot should be at a length sufficient to decode one bit and the sink cares only about the Boolean sum of the ``bit" patterns used by the transmitting sensors. Based on the filtered sequence (the energy detection sequence in the figure) the sink deduces which set of transmitted sequences could have generated the sequence received. In this example, there is no single sequence that can result in the received sequence; the only two sequences that can be combined into the received signal are the sequences corresponding to the second message of Sensor $2$ and the third message of Sensor $4$. Any other combination of sequences would result in a mismatch between the expected and the actual minislot energy levels. Note that even though in this toy example a simple Time Division Multiple Access (TDMA), which assigns each message a unique minislot can also work, in reality, where the overall number of sensors is high, such a solution \emph{requires many minislots (linear in the number of sensors) regardless of the actual number of sensors transmitting in each round}. Assuming the number of sensors expected to transmit is moderate, the solution we suggest in the sequel requires \emph{only a logarithmic number of minislots}, improving air time utilization and reducing the burden of contention for channel access and collision resolution procedures.

Another major challenge that our protocol can be easily modified to address is privacy and secrecy. In many such networks the information sent by the sensors is meant to be secure or private, i.e., we require that an eavesdropper (Eve) or an unintended addressee will not be able to decode the information, and sometimes not even identify the sender. Note that in the setup described above, since the sensors have limited information, identifying even the identity of the sending sensor may convey a lot of information, hence, in such cases, encryption is not very efficient. In this paper, we propose an enhancement to the suggested protocol, such that an eavesdropper receiving only parts of the channel output, is kept ignorant both of the content of the message and of the identity of the senders, at the expense of a small enlargement of the patterns allocated for each message, depending on the level of secrecy required.

In particular, the contributions of this work are as follows:
\begin{itemize}
\item We present a new MAC protocol, for data collection from dense wireless sensor networks, in which the sensors are expected to transmit only sporadically, and only a predefined amount of information (one out of a bank of possible messages per sensor). In the suggested protocol, the sink can collect up to $K$ reports simultaneously without any management or scheduling.

\item We present a downlink version of the protocol as well, suitable for data dissemination from a sink to a set of $K$ out of $N$ sensors simultaneously, without any management overhead, a predefined schedule or even a message notifying the relevant set of sensors.

\item To support the protocols, we provide a codebook construction with a very simple encoding and decoding procedure, such that not only the code is efficient but also the transmitted codewords are self-contained and do not require headers, trailers or sender identity.
    Moreover, we suggest effective decoding algorithms based on Column Matching \cite{chan2014non}.

\item \rev{We extend the single hop setup considered throughout the paper a to multi-hop setup in which a message needs to traverse multiple hops (pass through multiple relays) before reaching its designated sink.}

\item We present a secure version of the suggested uplink protocol, in which an eavesdropper receiving only part of the channel output cannot decode the messages or even identify the senders. Effective decoding algorithms are given as well.

\item We consider advance techniques for wireless transmission, in which the devices in the protocol suggested use OFDMA with BPSK modulator.

\item We present a rigorous analysis, proving the reliability of the protocols as well as establishing sufficiency bounds on the parameters, for both the unsecured and secured versions. Specifically, given the total number of sensors, the maximal number of different messages each sensor can have and the maximal number of simultaneous transmissions that the sink can decode, we provide bounds on the size of the codewords. We present numerical results which confirm our analytical results.
\end{itemize}
The structure of this work is as follows. In Section \ref{sec:model}, the system model is described. In Section \ref{related}, we summarize the related work. In Sections \ref{sec:related} and \ref{formulation} the protocol is described. Section \ref{LowerBound} includes the code construction and the decoding procedure. In Section \ref{efficient_algorithms}, an efficient decoding algorithm is described. In \Cref{noise}, we analyze the effects of a noisy channel and possible solutions. \rev{In Section \ref{sec:multi-hop}, the multi-hop setup considered}. In Section \ref{Dissemination}, an analogous data dissemination protocol is described. \ifsup\else Section \ref{StrongLowerBound} includes the code construction and the decoding procedures of the SWSN model.\fi In \Cref{Practical}, we consider an OFDMA extention. Section \ref{Sim} gives simulation results of the WSN and SWSN protocols, while \revised{Section \ref{sec:Imp} describes an implementation of the protocol on of-the-shelve, simple sensors.} Section \ref{conc} concludes the paper. \rev{\ifsup Appendix \ref{StrongLowerBound} includes the code construction and the decoding procedures of the SWSN model.}\fi
\vspace{-0.2cm}
\section{System Model}\label{sec:model}
We assume a dense wireless sensor network that harvests data from the area covered by the network. The network consists of multiple sinks (cluster heads) each collecting reports from a large set of sensors independently. Throughout the paper, we will focus on a single such cluster consisting of a large set of wireless sensors, denoted by $\mathcal{N}$, and a single sink. We denote by $N=|\mathcal{N}|$ the total number of sensors in the network (cluster).
We assume that all sensors are connected to the sink node, \emph{but only in the following limited sense.} First, beacons by the sink should be heard by all sensors. Then, in our upstream model (from the sensors to the sink) it is sufficient for the sink to be able to detect only whether there were transmissions (one or more) on the channel. For example, whether the received SNR is above a threshold. The sink \emph{does not necessarily need to decode any information from a single transmission from a single sensor}, and the received SNR is not necessarily above a decodable threshold. Specifically, we assume that in order to identify transmissions on the channel, the sink simply employs an energy detection mechanism (similar to traditional "carrier sense") in which it can recognize whether there is a transmission on the channel or not, yet it cannot identify the exact number of such simultaneous transmissions, and certainly cannot directly decode information from them.

We assume that each sensor $i$ has $C_i$ different messages it can transmit. There are no restrictions on the message length each sensor holds in its message bank. There are also no constraints on the message content each sensor has. For simplicity we will assume throughout the paper that $C_i = C, \forall i$. Nonetheless, an extension to different $C_i$'s is straightforward. We further assume that each sensor needs to transmit a message sporadically. In particular, we will assume that the sensors employ a duty cycle mechanism in which they randomly wake up and transmit a message unless they have an urgent message they need to transmit, in which case they wake up instantly waiting for transmission opportunity. We assume that the wakeup times including the urgent report instances are arranged such that the probability that more than $K$ sensors are awake at the same time waiting for transmission opportunity is very low. \emph{Note that we have no restriction on $K$ (i.e., $K\leq N$), however, we emphasize that the smaller $K$ is compared to $N$, the more efficient is the protocol we suggest}.

In the second part of the paper, we address a secure version, in which we assume that an eavesdropper is present, and observes a noisy version of what is received by the sink node. In particular, we assume that the eavesdropper has the exact same abilities as the sink including knowledge of all necessary codes, yet it can observe the channel only a $\delta$ fraction of the time; specifically the eavesdropper listens to (observes) the activity on the channel with probability $\delta$ and does not listen with probability $1-\delta$, e.g., if the time is slotted, the probability that the eavesdropper will observe or not observe the activity on the channel per slot is $\delta$ and $1-\delta$, respectively. Note that this model is similar to the common erasure channel (or packet erasure channel) but the erasures are on the channel activity and not on a bit (or a packet).

\vspace{-0.2cm}
\section{Related Works}\label{related}
We divide the discussion on related works into two parts. In the first, we provide a brief overview of several multipurpose WSN MAC protocols. Then, since our protocol is inspired by the classical Group Testing (GT) approach, in the second part we give a brief overview of related GT results.
\subsubsection*{WSN MAC protocols}
Since on the one hand, one of the foremost objectives of WSN is energy conservation, and on the other sensor nodes are expected to report only sporadically (and many of the reports can tolerate a short delay), most of the MAC protocols which were designed for WSN over the past decade and a half rely on a duty cycling technique in which each sensor node turns its radio on only periodically, alternating between active and sleeping modes \cite{polastre2004versatile,ye2002energy,huang2013evolution}. Such protocols took different approaches to address the rendezvous challenge in which a sender and a receiver should be awake at the same time in order to exchange information. In the synchronous approach, nodes' active and sleeping periods are aligned, i.e., all sensor nodes are active at the same time intervals and are required to contend for transmission opportunities during these intervals, e.g., \cite{lin2012sct, liu2016tas, kakria2014survey}. The asynchronous approach allows sensor nodes to choose individual wakeup times, maintaining unsynchronized duty-cycles, and employing various strategies to detect transmissions in the network and enable rendezvous between senders and receivers, e.g., \cite{buettner06xmac, sun2008ri,tang2011mac}. However, all these protocols are designed to support various types of traffic patterns, diverse topologies (e.g., single and multi-hop topologies) and most importantly, to support any kind of information exchange between the sensors. Accordingly, they have adopted the traditional approach in which the proposed channel access mechanism is independent from the message payload exchange between the sensor nodes, \emph{at the price of data encapsulation and signaling overhead}. In this work, since the information each sensor needs to convey is limited to one out of a number of known messages, we take a cross-layer design in which the coding and the channel access algorithm are intertwined. \rev{Indeed recently, data aggregation was used to compress the data and reduce its redundancy in order to save network energy before transmission to the sink \cite{tsai2013distributed,dhand2016data,boubiche2018big}. In a sense, in the protocol we give herein, using the coding we suggest, \textit{the channel inherently compresses the data efficiently} (based on a modified GT approach) by the addition in the air of the $K$ messages transmitted simultaneously from the sensors to the sink.}
\subsubsection*{Group Testing}
Classical group testing was used during World War II in order to identify syphilis infected draftees while dramatically reducing the number of required tests, by examining pooled tests of mixed blood samples \cite{dorfman1943detection}.

The concept was adopted later for multi access protocols. Specifically, MAC protocols adopted the GT philosophy for Collision Resolution Protocols (CRP). The basic idea behind these protocols is to resolve collisions whenever multiple users are trying to access the channel simultaneously, e.g., the binary-tree CR, the epoch mechanism or the Clipped Binary-Tree Protocol. An extensive survey of such protocols is given in \cite[Chapter 5]{rom1990multiple}.
However, all these protocols utilized the GT concept \emph{only as a collision resolution mechanism}, i.e., in case a collision occurred, they used the concept in order to decide who should contend for the channel next and when. Data was decoded successfully only when a single node transmitted without collision, using the standard layering and encapsulation. \revised{Thus, it is important to note that while there are indeed many works in the literature which utilize GT concepts in communication networks and other models \cite{du1999combinatorial,macula1999probabilistic,wu2015partition,wu2015detection,malloy2014near,de2017epsilon,aksoylar2017sparse,tan2014strong}, none suggest a protocol for collecting or disseminating data in wireless sensor networks.}
The main contribution of this paper is in suggesting a protocol which decodes all data, from all simultaneously transmitting sensors, using a novel extension to this concept, namely, analysing the location of the colliding and non-colliding minislots in order to identify \textit{both the senders and the data sent}.
%
\section{WSN Data Collecting Protocol Design}\label{sec:related}
\begin{figure}
  \centering
  \includegraphics[trim=0cm 0cm 0cm 0cm,clip,scale=0.3]{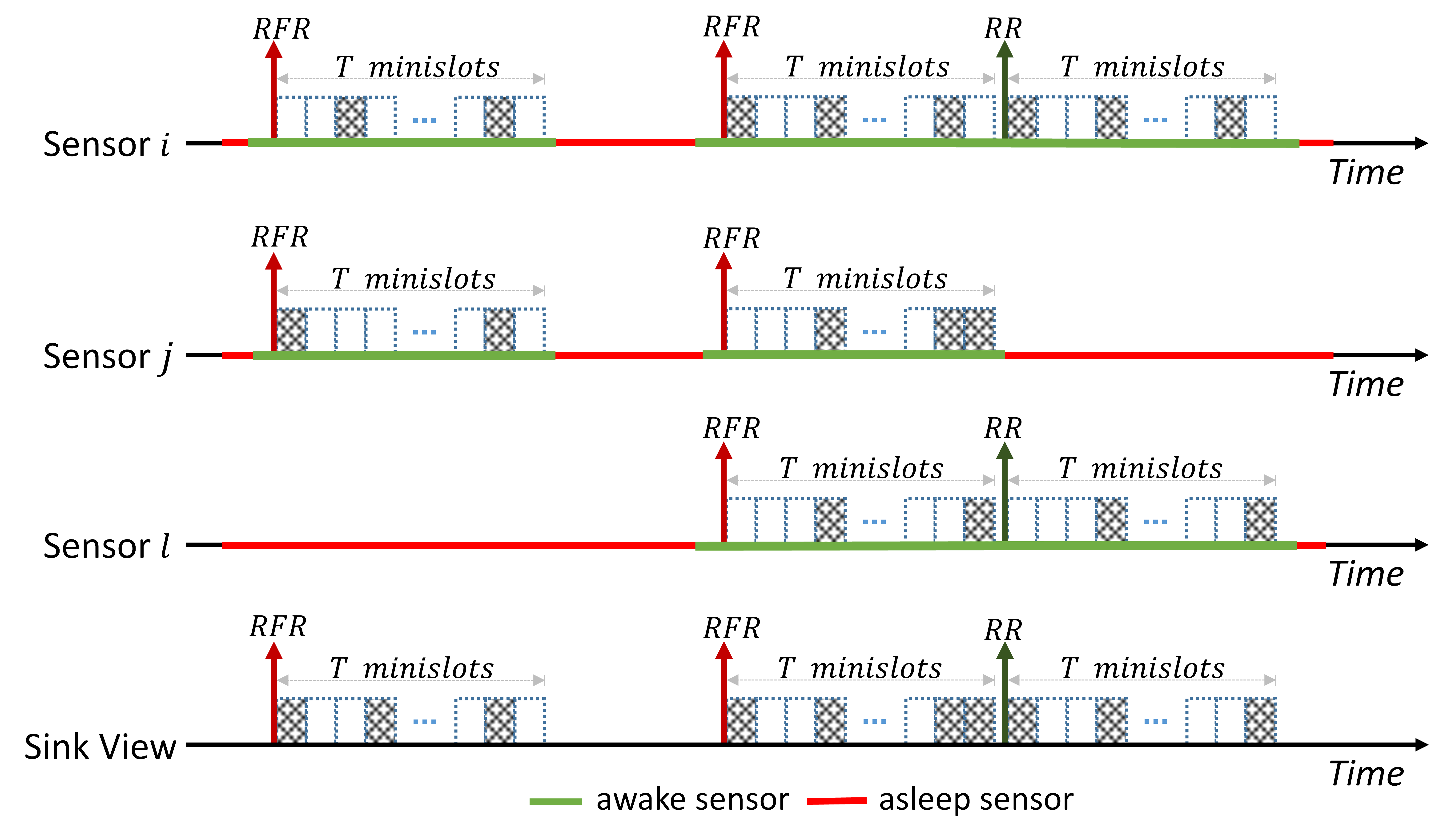}
  \caption{Basic protocol operation.}
  \label{fig:Overview}
  \ifshort \vspace{-4mm} \fi
\end{figure}
In the suggested protocol, the sink periodically transmits a predefined beacon, which starts a report transmission interval. We term this beacon RFR (Request For Reports). The RFR is then followed by a sequence of $T$ minislots. Figure~\ref{fig:Overview} provides an illustration of the protocol operation. Note that there is no need for synchronization or for each sensor to keep track of the minislot boundaries at all times, as the awake sensors waiting for transmission can synchronize based on the received RFR. Denote by $\tau $ the maximum propagation delay between any sensor and the sink, and by $\eta$ the time required by the sink to identify that there is transmission going on. Note that $\eta$ can be very short, as it only corresponds to the duration required by the sink to sample the channel and identify that there is a transmission going on (there are no headers, preambles or data involved). We assume that the minislot duration is longer than $2\tau + \eta$, which is sufficient for all sensors to receive the RFR, start a transmission of duration $\eta$ and for the sink to receive all the transmissions starting in this minislot. I.e., as far as the sink is concerned, the time duration of a minislot is such that there is no transmission that can start on the current slot and leak to the next slot. It is reasonable to assume that $\eta \geq \tau$, hence the minislot duration is greater than $3\tau$. After transmitting the RFR, the sink node switches to receive mode and identifies whether there was a transmission on each of the following $T$ minislots. Recall that the sink detection is binary, i.e., it can only recognize whether there was a transmission in a minislot or not. It does not try to detect how many sensors transmitted during an occupied minislot.

Each sensor is assigned a unique sequence of ones and zeroes of length $T$, for each of its messages. The construction of the sequences is given in Section~\ref{LowerBound}. A sensor wishing to send a report wakes up and waits for the RFR. After receiving the RFR, the sensor follows the sequence associated with the message it wishes to transmit, transmitting "energy" of duration $\eta$ (in the form of a pre-defined signal) at each minislot in which the corresponding bit in the sequence is one.

After the $T$ minislots interval, the sink has a sequence of ones and zeroes of length $T$, indicating on which of the minislots it identified transmission. We denote by $Y(t)$ the sequence observed by the sink at the $t$-th interval. Based on the observed $Y(t)$ the sink tries to decode the messages transmitted by the sensors (we provide two decoding algorithms in Sections~\ref{LowerBound} and \ref{efficient_algorithms}). Note that since the sequences are unique, each sequence indicates the identity of the sender and the message sent. If the sink is not able to decode the received sequence $Y(t)$, which, as we prove in Section~\ref{LowerBound}, can only happen if the number of transmitting sensors in the interval was greater than the expected number $K$ for which the sequences were designed, it transmits another beacon, termed Retransmission Request (RR), which starts the exact same procedure as the RFR only this time sensors waiting for transmission participate in the following interval only with some probability. The probability for participating in the following interval is predefined and can prioritize different messages (e.g., messages with high urgency will receive high probability and messages that can tolerate delay will be assigned low probabilities and sensors with non-urgent messages can go back to sleep waiting for their next wakeup time). The discussion on the collision resolution probabilities is beyond the scope of this paper, as our novelty lies in the transmission protocol and its ability to allow multiple transmission of messages without the various MAC and upper layers overheads. The collision resolution procedure is rather standard and can be repeated multiple times.

It is important to note that the suggested protocol can be \emph{interleaved within traditional wireless sensor protocols}, in which some or all sensors are required to send occasionally a regular report. For example, the suggested protocol can be incorporated within the operation of RI-MAC \cite{sun2008ri}, such that occasionally the sink transmits an ordinary RI-MAC beacon, which is different from the RFR beacon, to initiate an RI-MAC operation interval, i.e., the RI-MAC beacon will be followed by ordinary DATA transmissions according to the ordinary RI-MAC protocol. In the same manner, the suggested protocol can be combined with transmitter initiating protocol such as X-MAC \cite{buettner06xmac}, such that a sensor wishing to report a typical DATA packet transmits a sequence of short preambles prior to DATA transmission according to the X-MAC protocol.

\rev{Moreover, in Section~\ref{sec:multi-hop}, we extend the single hop setup presented in this section, to support a multi-hop setup in which a message needs to traverse multiple hops before reaching its designated sink.}
\subsubsection*{\rev{Power Consumption}}
\rev{energy efficiency is an important performance criterion in wireless sensor networks. The protocol suggested in this paper is a MAC protocol in which the receiver initiates the data transmission. It has been shown that such protocols are highly energy-efficient over a wide range of traffic loads and topologies (e.g., \cite{sun2008ri, tang2011pw, tang2011mac}). However, such protocols are prone to collisions when the network is dense and multiple devices are waiting to transmit simultaneously to the same receiver (sink, or a relay in a multihop topology). Such collisions are energy wasteful as they not only waste the energy consumed by the devices in the failed transmission but they also require a collision resolution mechanism to allow retransmission of the lost packets, a mechanism which is typically consume high energy. The protocol introduced in this paper improves energy consumption over the existing receiver initiated MAC protocol by four means. First and foremost, it allows the transmission of up to $K$ devices simultaneously, hence dramatically reduces the collision probability. Note that $K$ is a parameter chosen by design hence can be adjusted based on the topology and the expected traffic pattern. Second, the overhead involved in a packet transmission (the protocol data unit (PDU)) is dramatically reduced, hence transmission time is reduced. Third, the protocol relies on an energy detection mechanism which is robust to noise and interference hence suffers minimum transmission failures. Furthermore, simple energy detection requires a relatively low SNR to operate robustly. Fourth, the hardware requires only a simple energy detection mechanism with no need to run sophisticated algorithms, hence is expected to be very energy efficient.}

    %
\section{Model Formulation and Transmission Process}\label{formulation}
In this section, we formalize the \emph{WSN} model that is used throughout the paper. We denote the set of wireless sensors by $\mathcal{N}$. We will concentrate on a time instance right after an \emph{RFR} has been sent by the sink and a subset of unknown sensors comply with the \emph{RFR}, and transmit their reports. We denote this unknown subsets by $\mathcal{K}$. We denote by $N=|\mathcal{N}|$ and $K=|\mathcal{K}|$ the total number of sensors, and the number of active sensors at the same time slot, respectively. In the analytical part of the paper we will assume that the number $K$ of active sensors is known a-priori. The algorithms and results presented hereby can be easily adopted to the case where only an upper bound on $K$ is known, and the actual number is smaller. The case where more than $K$ transmit was briefly described at the end of Section~\ref{sec:related}. The sink objective is to determine which subset of the sensors were active and what is the information (messages) they transmitted. Throughout the paper, logarithms are in base $2$.

Each of the sensors has its own set of $C$ independent messages. We denote each such message by $M_{n,c}$, $n\in\mathcal{N}, c\in\{1,\ldots,C\}$, and sensor $n\in\mathcal{N}$ message list by:
\ifshort
\begin{equation*}
\textstyle \textbf{M}_n=[M_{n,1};M_{n,2};\ldots; M_{n,C}]
\end{equation*}
\fi
We further consider all the possible sets of $K$ active sensors, and denote by $W \in \{1,\ldots, {N \choose K}\}$ the index of the \emph{subset $S\subset \{1,\ldots,N\}$ of sensors active and transmitting at the same time}. Thus, $S_w$ denotes the $w$-th subset of size $K$ out of the $N$ sensors.
\off{We assume that $W$ is uniformly distributed, that is, there is no \emph{a-priori} bias to any specific subset of active sensors.}
\rev{The WSN solution we consider is fair in a sense that if sensors have the same transmission probability, they have the same probability to access the channel and the protocol suggested dose not prefer sensors over others.} We further denote by
\ifshort
\[
\textstyle \textbf{M}_{W}=[M_{w_1,c_1};M_{w_2,c_2};\ldots; M_{w_K,c_K}]
\]
\fi
the $K$ messages transmitted by the active sensors (members of $S_w$) to the sink. Note that each row in $\textstyle\textbf{M}_{W}$ corresponds to a separate message, that is, $M_{w_i,c_i}$ is the $c_i$th message of the $w_i$th sensor in the active set.
\begin{figure}
  \centering
  \includegraphics[trim= 0.5cm 0cm 0cm 0cm,clip,scale=0.95]{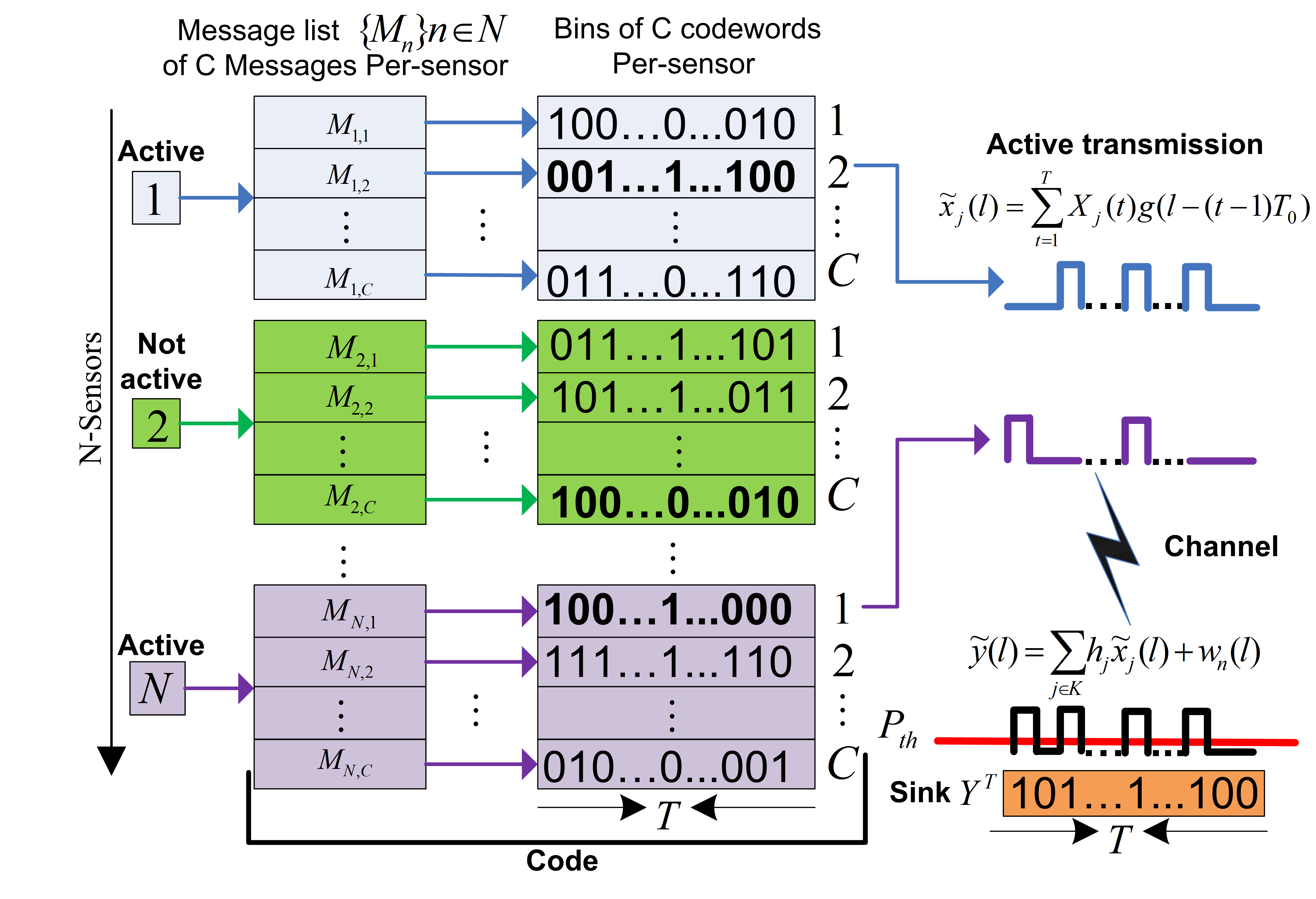}
  \caption{Encoding, transmission and detection in the suggested protocol.}
  \label{fig:WiretapCoding}
  \ifshort \vspace{-4mm} \fi
\end{figure}

As previously described, for each message a sensor has a unique sequence, which in the sequel we refer to as codeword. That codeword is the one transmitted whenever the sensor intends to send this message. In other words, if sensor $n$ is set to send $M_{n,c}$, it uses the codeword $X^{T}_{n}$ associated with this message. Each message has a different codeword associated with it (we drop the message index for clarity). Given the particular messages sensors intend to transmit, we define a \textit{transmission} matrix
\ifshort
\begin{equation*}
\textstyle \textbf{X}=[X_{1}^{T};X_{2}^{T};\ldots; X_{N}^{T}] \in \{0,1\}^{N\times T},
\end{equation*}
\fi
where each row corresponds to a codeword, describing the message a sensor may transmit \emph{if it is active}.

Assuming that sensors use Power Amplitude Modulation (PAM), we denote by
\[
\tilde{x}_{j}(l)=\sum_{t=1}^{T} X_{j}(t)g(l-(t-1)T_0), \text{ }\text{ } 0 \leq l \leq T \cdot T_0,
\]
the signal transmitted by the $j$-th sensor, where minislot $t$ is defined by $(t-1)T_0 \leq l \leq tT_0$ and $g(l), 0 \leq l \leq T_0$ denotes the PAM pulse.
The channel output signal $\tilde{y}(l)$ is given by
\[
\tilde{y}(l)=\sum_{j\in \mathcal{K}} h_j \tilde{x}_j(l) + w_n(l),
\]
where $h_j$ is a channel fade for the $j$th active sensor and $w_n$ is an additive noise at the sink. Note that the fade herein is fixed for the entire transmission only for simplicity of exposition, and it may depend on $t$ as well. Figure \ref{fig:WiretapCoding} depicts an example.

We denote by $P_{th}$ the power threshold of the sink's hard decision mechanism. Hence, the outcome vector $Y_T =(Y(1), \ldots, Y(T))$ at the sink is binary, with $1$ in a minislot $t$ if
\[
\int_{(t-1)T_0}^{tT_0} \tilde{y}(l)g(l-(t-1)T_0)dl\geq P_{th},
\]
 and $0$ otherwise.
In the first part of this work we assume that the noise in the channel cannot produce errors at the sink.
In \Cref{noise}, we will extend the model, and consider the case in which noise can produce positive and negative errors.	

We denote by $\hat{W}$ and $\hat{\textbf{M}}_{\hat{W}}(Y^T)$ the set of sensors estimated by the sink to be the transmitting set, and the estimated set of messages sent by them, respectively, according to the received signal $Y^T$.
We refer to the possible transmission matrix, together with the decoder as a \emph{WSN algorithm}. The following definition lays out the goals of the WSN algorithm.
\begin{definition}\label{def:reliable_secure}
A sequence of WSN algorithms with parameters $N,K,C$ and $T$ is asymptotically \emph{reliable} if at the sink, observing $Y^T$, we have
\[
\lim_{N \to \infty} P(\hat{\textbf{M}}_{\hat{W}}(Y^T) \ne \textbf{M}_W) = 0,
\]
i.e., the error probability both in the active set and the message associated with this set, goes to zero as the number of sensors goes to infinity.
\end{definition}

In \Cref{Sim}, we show that our coding and decoding procedures are successful at moderate and even small values of $N$. In the next section, we construct a code (for parameters $N$, $K$ and $C$) which associates a codeword of length $T$ to each of the $N \cdot C$ messages ($C$ per sensor) and a decoding algorithm $\hat{\textbf{M}}_{\hat{W}}(Y^T)$, such that observing $Y^{T}$, the sink will identify the subset of active sensors and the messages transmitted by them with desired high probability.       %
\section{Code Construction and Decoding at the Sink } \label{LowerBound}
We now turn to the construction and decoding in detail.
\subsection{Codebook Generation}
For each sensor $n\in \mathcal{N} $, we map each message $c\in \{1,\ldots,C\}$ to a codeword.
We now provide the construction of a codebook per sensor.
Let $P(x)\sim Bernoulli(\ln(2)/K)$.
Using a distribution \ifshort$P(X^{T})=\prod^{T}_{i=1}P(x_i)$, \fi
generate $C$ independent and identically distributed codewords $x^{T}(c)$, $1 \leq c\leq C$.
We denote this set of codewords as the $n$-th "bin". Thus, a bin contains $C$ codewords of size $T$. Bins across sensors are independent as well.
The codebook is depicted in the central side of Figure \ref{fig:WiretapCoding}. Reveal the codebooks to the sensors and the sink.
\subsection{Decoding at the Sink}\label{LowerBoundDec}
The first decoder we suggest is the optimal decoder, Maximum Likelihood (ML).
This decoder will declare the right set of $K$ messages transmitted (and the active sensors) with a high probability if $T$ satisfies the bound \Cref{direct lemma1} below.
\rev{However, such a decoding algorithm is complex and assume that $W$ is uniformly distributed, that is, there is no \emph{a-priori} bias to any specific subset of active sensors. Hence, in \Cref{efficient_algorithms}, we consider a computationally efficient algorithm without any assumption required on $W$ distribution.}

As described in \Cref{formulation}, in the first decoding step, the sink uses a hard decision mechanism to achieve the binary channel output vector $Y^T$. After $Y^T$ is obtained, the ML decoder looks for a collection of $K$ codewords $\hat{\textbf{X}}_{S_{\hat{w}}}^{T}$, \textit{each one taken from a separate bin}, for which $Y^T$ is most likely. Namely,
\ifshort
\[
\textstyle P(Y^{T}|\hat{\textbf{X}}_{S_{\hat{w}}}^{T})>P(Y^{T}|\textbf{X}_{S_{\grave{w}}}^{T}), \forall \grave{w} \neq \hat{w}.
\]
\fi
That is, the sink looks for both the set $\hat{w}$, and the messages $\hat{c}_j, j\in S_{\hat{w}}$ which explain $Y^T$ the best.
Then, the sink declares $\hat{W}$ as the set of active sensor and maps the rows in $\hat{\textbf{X}}^{T}_{S_{\hat{w}}}$ back to the messages, as this is a $1 : 1$ mapping.
\subsection{Reliability}
The following lemma is a key step in proving the reliability of the decoding algorithm.
\begin{lemma}\label{direct lemma1}
If the size of the codewords satisfies
\begin{eqnarray}\label{eq:reduce_hw}
 T \ge \max_{1 \leq i \leq K}\frac{(1+\varepsilon)K}{i}\log\binom{N-K}{i}C^i,
\end{eqnarray}
then, under the codebook above, as $N\rightarrow \infty$ the average error probability approaches zero.
\end{lemma}
Note that using the upper bound $\log\binom{N-K}{i} \leq i\log \frac{(N-K)e}{i}$, the maximum in \Cref{direct lemma1} is easily solved, and we have
\[
T \geq  (1+\varepsilon) K \log (N-K)Ce.
\]
That is, assuming a WSN algorithm for the protocol described in Section \ref{formulation}, with the parameters $N$, $K$ and $C$,
if the size of the codewords is
\ifshort
\begin{equation}\label{order_T}
\textstyle T = \Theta \left(K \log N C\right),
\end{equation}
\fi
observing $Y^{T}$ at the sink, the decoder can identify the sensors and the messages with high probability.
\begin{corollary}
  Assume each sensor has $B$ bits to send. That is, $B=\log C$. Then the length, in single-bit minislot, required to send $B$ bits from $K$ sensors, out of $N$, without knowing in a once which will transmit and without setting any pre-defined schedule is  $\Theta \left(K \log N +KB\right)$, namely, besides the obvious linear dependence in $KB$, there is only logarithmic added factor in $N$.
\end{corollary}

The proof of \Cref{direct lemma1} extends the results given in \cite[Theorem III.1]{atia2012boolean} to the codebook required for WSN. Specifically, we may interpret the analysis in \cite[Section III]{atia2012boolean} as analogous to the case where each sensor has \textit{only one message in its bin}. However, in the WSN protocol suggested herein, each sensor has $C$ messages in its bin, and the decoder has $\binom{N}{K}C^K$ possible subsets of codewords to choose from, $\binom{N}{K}$ for the number of possible sensors and $C^K$ for the number of possible rows (messages) to take in each bin. The analysis is complex, taking into account the fact that rows have to be selected from different bins, and is out of the scope of this version. In the technical report completing this submission, \ifsup Appendix \ref{appendix:APPENDIX A} \else \Cref{appendix:APPENDIX A}\fi, we formally analyze the bound on the error probability, based on our earlier work \cite[Lemma 2]{sgt2016arxiv}.
        %
\section{Efficient Decoding Algorithms}\label{efficient_algorithms}
In the decoding algorithm described in Section \ref{LowerBound}, as well as the secured version given in \ifsup Appendix \ref{StrongLowerBound} \else \Cref{StrongLowerBound} \fi, the sink uses an ML decoder to identify the sub-set of $K$ active sensors and the messages they transmitted. Such a decoding algorithm may suffer from high computational complexity, especially when the number of simultaneously transmitting sensors, $K$, is high. In this section, we present a computationally efficient algorithm. Specifically, we harness the Column Matching (CoMa) procedure \cite{chan2014non}, which utilizes a much simpler decoding rule at the price of only a slightly lower rate, i.e., slightly higher codeword length $T$. The resulting decoding process at the sink is very efficient, and can be implemented for practically any $N$ and $K$.

CoMa relies on the observation that a busy minislot (i.e., a minislot for which the sink detected energy on the channel) can only indicate that at least one of the sensors transmitted in this minislot, yet, using a busy minislot one cannot validate or invalidate any \emph{specific codeword}. In other words, during a busy minislot, regardless of whether a corresponding bit in a codeword is 1 or 0, if the codeword was suspected as a transmitted codeword it remains as such. On the other hand, an idle minislot (i.e., a minislot in which no transmission was detected by the sink) \emph{can invalidate codewords in which the corresponding bit was 1}, i.e., all codewords for which the corresponding bit is 1 could not have been transmitted (disregarding noise; The noisy case is discussed in \Cref{noise}). Accordingly, CoMa examines all minislots for which no transmission was detected, and invalidates the codewords that 1 appears in the corresponding bit.

\begin{figure}
\centering
\includegraphics[trim= 0cm 0cm 0cm 0cm,clip,scale=0.3]{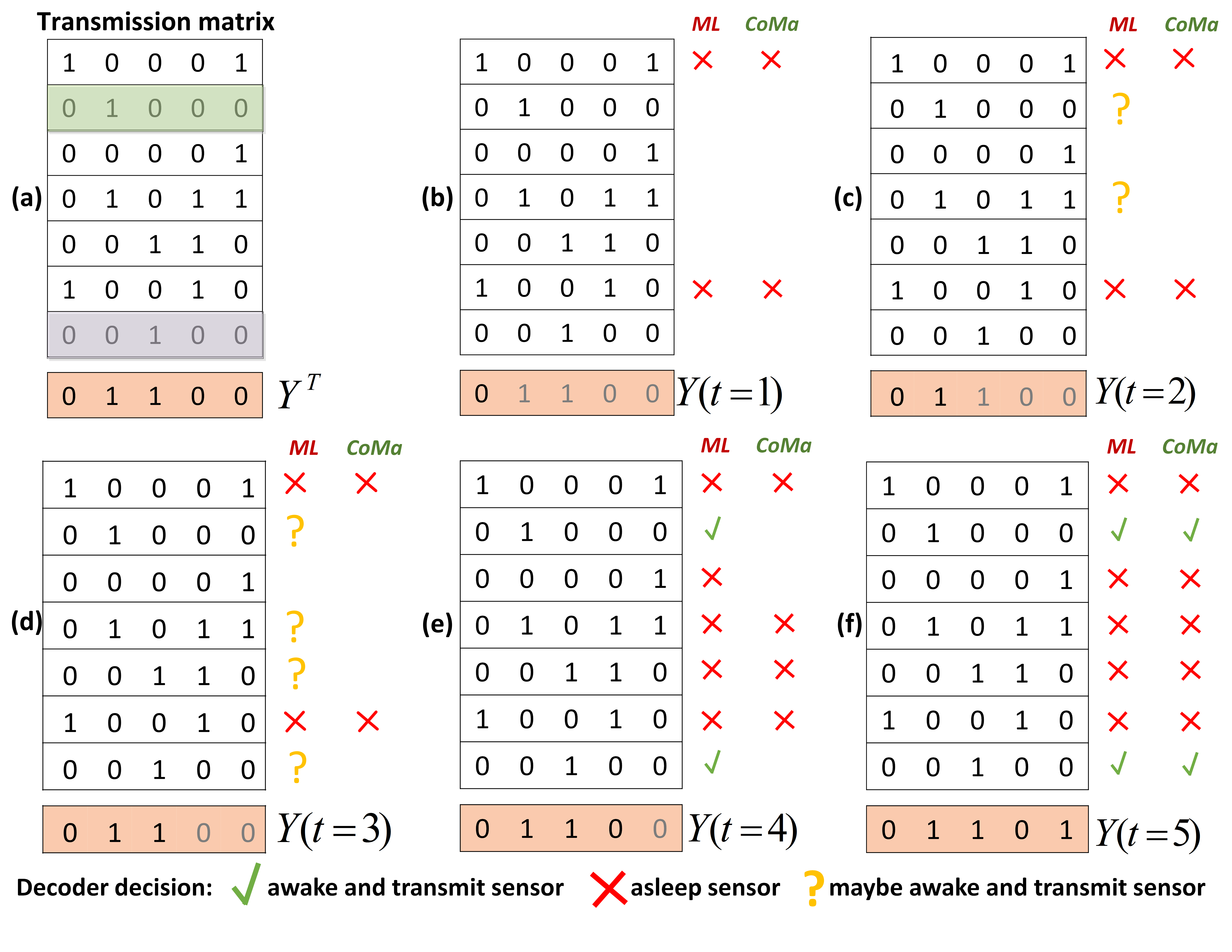}
\caption[]{An example of decoding at the sink using ML and CoMa methods.}
\label{basic_testing}
\ifshort \vspace{-4mm} \fi
\end{figure}

Figure \ref{basic_testing} depicts a small example of the efficient decoding algorithm suggested compared to ML decoding. This example includes 7 sensors ($N=7$); each has only one possible codeword to transmit ($C=1$) and the expected number of simultaneously transmitting sensors is 2 ($K=2$).
\Cref{basic_testing} (a) presents the codewords matrix such that row $i$ corresponds to sensor $i$'s codeword. In the example the second and seventh sensors are awake and transmit their codewords, while all other sensors are asleep. Accordingly, the last row indicates the channel output after energy detection, $Y^T$. It is important to note that since both the code construction as well as the channel are memoryless across minislots, both the suggested algorithm and ML can be easily viewed on a minislot-by-minislot basis (per-column).

Examining the first column in \Cref{basic_testing} (b), since the output in the first minislot indicates that there was no transmission in this minislot ($Y(1)=0$), the sink can conclude by both of the decoding methods that sensors 1 and 6 are not the transmitters. In the second minislot (\Cref{basic_testing} (c)), the output indicates that there was a transmission ($Y(2)=1$). The CoMa decoder disregards this minislot. The ML decoder, however, is stronger and may go over all pairs of codewords. It can thus infer (based on this minislot alone) that the pair of sensors which transmitted must include at least one of the sensors 2 or 4. In the same manner, since $Y(3)=1$ in \Cref{basic_testing} (d), CoMa does not gain any new knowledge, yet the ML decoder now infers that at least one of sensors 5 and 7 were transmitting as well. In fact, the ML decoder has more knowledge. It can now know that while 2 and 4 were possible transmitters before, now they cannot be the transmitting \emph{pair} (this is not depicted in the figure).
Since $Y(4)$ indicates that there was no transmission in the 4th minislot, both decoders can conclude that sensors 4, 5 and 6 are not possible transmitters. Accordingly, the ML decoder can deduce that sensors 2 and 7 are the transmitting sensors and finish the decoding process. However, the CoMa decoder, which invalidated only 4 sensors (sensors 1, 4, 5,and 6) still has 3 legitimate candidates for transmission (\Cref{basic_testing} (e)).
Based on the 5th minislot, which indicates that there was no transmission at that time, CoMa can invalidate sensor 3 and reach the same conclusion as ML.

Section~\ref{Sim} presents numerical results which validate the above intuition, and depict the actual message lengths required by both algorithms.
     %
\section{Effects of a Noisy Channel}\label{noise}
One of the advantages of the suggested protocol is its robustness to noise and interference. Specifically, in the protocol suggested, the sink node only relies on a simple energy detection mechanism in which on each minislot it just needs to detect whether there was a legitimate transmissions (one or more), or not, i.e., similar to the traditional carrier sense mechanism, whether the received signal is above the noise floor or not.

While such a mechanism trades off rate for robustness very well, errors may occur due to strong interference from unregistered devices or due to sporadic high noise. There are two possible errors: false positive detection, which is more common, in which the sink detects energy despite the fact that no legitimate sensor was emitting energy in the minislot, and the less common one, false negative, in which the sink does not detect energy despite the fact that at least one legitimate sensor had emitted energy during the minislot. The latter type can happen due to temporary channel changes (e.g., deep fades). Note that at without explicit error correction, at least for the first type (the more common), an error may result only in false positive detection of a message, i.e., the sink will identify an additional message that was not transmitted, yet there will be no false negatives for which the sink will lose a legitimate transmission (there can be an undecodable reception in which a retransmission is required). Yet, in the sequel, we describe an error correction coding scheme, at the price of only a slightly higher codeword length $T$, depending on the noise level it needs to overcome.

As for the optimal ML decoder suggested in \Cref{LowerBound}, the enhancement is straightforward. Since the decoder looks for a set of codewords, for which the signal received at the sink is the most likely set that has generated this outcome, regardless of whether there is a perfect match between the  combined set and the outcome on all the minislots, all the decoder needs to do is add sufficient redundant minislots to ensure that the genuine set is discovered despite the few erroneous minislots. The full analysis of the number of minislots required to add using ML decoding follows from \cite[Section VI]{atia2012boolean}, with the consideration that each sensor has at most $C$ possible messages, similar to our analysis in \Cref{LowerBound} and \ifsup Appendix \ref{appendix:APPENDIX A} \else \Cref{appendix:APPENDIX A} \fi for the error-free case.

Overcoming the noisy channel using CoMa is a bit more challenging. Recall that CoMa relies on the fact that a received minislot with below the threshold energy, indicates that all the possible sequences with $1$ in this specific minislot can be eliminated, as they were definitely not in the generating set. However, with a noisy channel such a deduction cannot be asserted, as each such minislot can be erroneous. Accordingly, we adopt the Noisy-CoMa suggested for the non-secure GT \cite{chan2014non}. Specifically, we relax the firm CoMa requirement that each and every $1$ in a transmission sequence must overlap with a high energy minislot in the additive outcome. Instead, we allow for a certain number of ``mismatches", which is determined based on both the number of ones in each transmission sequence and on the noise level we wish to be protected from (specifically, on the probability $q$ that the energy detector will provide an erroneous decision, interpreting high energy minislot as low energy, or vice versa).

Formally, for a codeword $c$ of sensor $n$, we define by $\zeta_{n,c}$ its support, i.e., the set of indices which correspond to $1$ in the sequence ($\{t:\textbf{X}_{n,c}(t) = 1\}$). We also define the matching set $\beta_{n,c}$ as the intersection of $\zeta_{n,c}$ and the support of $Y$, that is, where both $Y(t) = 1$ and $\textbf{X}_{n,c}(t) = 1$. The decoder the uses a ``relaxed'' rule, in which it allows a codeword to have less than a fraction $q$ of minislots in which the codeword has $1$ but the received signal was detected as below the threshold signal ($|\beta_{n,c}| \geq |\zeta_{n,c}|(1-q(1+\epsilon))$, for $\epsilon >0$). All codewords with a fraction of mismatches less than $q$ are declared as transmitted. This procedure overcomes the errors occurring by the energy detection mechanism, at the price of a slightly higher $T$ compared to the original CoMa algorithm given in \Cref{efficient_algorithms}.
                         %
\section{\rev{Multi-hop paths}}\label{sec:multi-hop}
\rev{Throughout this paper we mainly concentrate on a single hop path setup in which each device (sensor) is a single hop away from its respective sink, i.e., the message transmitted by a sensor is directly received by its associated sink without the need for any relay. However, the protocol presented herein can be easily modified to support multi-hop paths in which a message needs to traverse multiple hops (pass through multiple relays) before reaching its designated sink. In this section we suggest possible modifications which support such multi-hop paths setup.}

\rev{The enhancements suggested herein rely on hierarchical or cluster-based routing which is widely studied in the context of data forwarding in wireless sensor networks. In cluster-based routing, all sensors are organized in groups termed clusters. Each cluster includes a Cluster Head (CH) which collects the messages from its own cluster members and forwards it to its designated sink. The information (messages) can traverse several relays before reaching the sink. These relays, that forward the traffic toward the sinks, can be dedicated entities, other CHs or other sensors. We distinguish between the tier in which each CH collects the information from all its devices and the tier that inter-connects the CHs to their designated sinks; we denote the latter the backhaul tier. Several hierarchical routing protocols have been developed based on such cluster-based-architecture, and it has been shown that such protocols not only address the multi-hop routing challenge, but also enhance many other performance objectives such as energy consumption, load balancing and the scalability problem of large WSNs (e.g., \cite{singh2015survey,akyildiz2002wireless}). In the sequel we present two approaches to extend the protocol suggested in this paper to support multi-hop traffic. We assume that the first phases in which clusters are formed, CHs are selected and a routing protocol discovering the routes between each CH and its designated sink are performed based on any preferable protocol (e.g., \cite{singh2015survey,akyildiz2002wireless}).}

\rev{We suggest two approaches to forward the messages along the routing paths toward the set of sinks. In both approaches each CH collects the messages from its designated sensors based on the suggested protocol (Section~\ref{sec:related}). The difference is in the forwarding tier (the backhaul tier). Specifically, in the first approach the forwarding mechanism that forwards the messages collected by each CH toward the sink is performed via any traditional MAC protocol (e.g., \cite{sun2008ri,buettner06xmac,akyildiz2002wireless}). Recall that as discussed in Section~\ref{LowerBound} the suggested protocol can be interleaved within traditional wireless sensor protocols. Accordingly, the data collection and the forwarding procedures can utilize different protocols. Note that relying on this approach, each relay does not need to decode the messages or identify the senders and can simply forward the received sequence ($Y(t)$) as the message payload. Accordingly, the relays need not hold the code books or perform any decoding procedure. Each sink will hold the code-books of its designated devices and perform the decoding procedure for each cluster separately, as described in Section~\ref{LowerBound}. Further note that the length of the sequences assigned to each device per message ($T$) which reflects on the number of transmission mini-slots and eventually on the message length forwarded on the backhaul tier, can vary between the clusters and should be according to the number of devices in each cluster. Each sink should maintain the codebook for each of its designated clusters. Based on the received packet origin (the originating CH) and the respective codebook it can decode the messages.  On the other hand, the forwarding protocol on the backhaul tier is not only inheriting the overhead required by the selected protocol (headers, trailers, control fields, contention mechanism, etc.) but also is prone to the detriments characterizing the selected protocol (e.g., collisions, channel access delays, reliability).}

\rev{In the second approach both the data collection protocol and the forwarding procedure rely on the protocol suggested in this paper. The challenge is that different relays forward different numbers of messages from different sets of devices. Note that even in the same forwarding subtree the closer one gets to the sink the larger the number of messages that can traverse the relay and the larger the number of messages comprising the transmitted sequence. We suggest two different ways to resolve this challenge, each with attendant pros and cons.}

\rev{(i) \emph{Combine and Forward}: According to this scheme each relay needs to combine the received sequences to a single one and transmit it forward. This approach utilizes a single codebook for each subtree (each sink and all its descending devices share a single codebook). Accordingly, the length of the sequences assigned to each device per message ($T$) is the same for all messages destined to the same sink. The codebook for each such subtree is designed exactly as described in Section~\ref{LowerBound}, where the length of the codewords is determined according to the number of possible messages destined to each sink ($N,C$ and $K$ are determined according to the number of devices in the tree rooted at the sink, the number of expected messages per device in the tree and the estimated upper bound on the expected number of awakened devices simultaneously, respectively). Note that different sinks can have different codebooks with different codeword length and that the codebooks need not be coordinated between the trees rooted at different sinks. Note that as in the first approach, also in this case the relays do not need to decode the messages or identify the senders, instead whenever a relay needs to forward reports (receives the RFR indicating a new transmission cycle start) it just needs to combine all the received sequences (conduct Boolean sum of the “bit” pattern received from all its successors) and transmit the result sequence. Obviously, the cost is that all the transmission periods by all relays along the path to the sink are according to the total number of devices in the respective subtree.}

\rev{(ii) \emph{Encode-Decode-Combine and Forward}: Based on this scheme, each relay needs to encode the received sequence, re-decode the received messages based on a different codebook, combine the received sequences to a single one (conduct Boolean sum of the new decoded sequences) and transmit it forward. Accordingly, each relay will have its own codebook which needs to be coordinated only with its direct descendants. The length of the sequences (codewords) transmitted by the relay’s descendants is according to the number of possible messages from all its descendant devices (number of possible messages in the subtree rooted at the relay). Note that the transmission period grows the closer one gets to the sink, reaching its maximum on the last hop which connects the sink to its descendants (exactly the same as the length of the sequences at the Combine and Forward scheme). However, note that the cost is that each relay needs to decode the messages received, map each one to its corresponding longer sequence, locally combine them to the longer combined sequence (performing Boolean sum, emulating the transmission of the sequences) and transmit the newly generated sequence after receiving the RFR.}                  %
\section{Data Dissemination} \label{Dissemination}
Heretofore, we described and analyzed an efficient data collection protocol for WSN. In this section, we leverage the suggested method to design a data dissemination protocol, i.e., from a single source node to all sensors. We show that the advantages obtained for data collection also apply to the case of data dissemination. Similar to the upstream protocol (data collection between the sensors and the sink) the downstream protocol does not require any synchronization, coordination or management overhead.

The system model is basically the same as the one presented in Section~\ref{sec:model}. We assume that the sink is required to disseminate messages to the sensors. We assume that each sensor can receive up to $C_i$ messages, and that the sink is transmitting to at most $K$ sensors simultaneously. If the sink needs to transmit to more than $K$ sensors, it is required to break the designated sensors to groups of at most $K$ sensors each. Note that similar to the upstream case, $K$ is a design decision. Enlarging $K$ will enable the transmission to a larger group simultaneously, yet will consume more overhead per transmission. Similar to the upstream case, we assume that the sensors are quite limited, and are only required to employ a simple energy detection mechanism in which they just need to recognize whether there is a transmission on the channel or not (i.e., whether the received SNR is above a threshold or not).

Next we describe the design of the downstream protocol. After providing an overview of the protocol, we describe the encoding procedure at the sink and the decoding procedure at each sensor. Similar to the upstream case, and utilizing the same coding techniques, the downstream case supports unsecured as well as secured versions.
\subsection{Downstream Protocol Design}
In the downstream case, the Sink node is required to periodically send a beacon. This beacon can take two forms, if the Sink is not required to send any data to any of the sensors, it sends a predefined Null beacon (denoted as NB). However, whenever the sink is required to send a message to one or more sensors (but no more than $K$), it transmits a predefined Data Dissemination beacon (denoted as DDB), which initiates a sequence of $R$ minislots (similar to the $T$ minislot epoch presented in Section~\ref{sec:related} for the data collection protocol). During this dissemination interval, the sink transmits its encoded message to all the designated sensors in the bunch, according to the encoding scheme presented below. Note that similar to the upstream case neither the DDB nor the transmitted sequence need to contain the identity of the addressee sensors.
\begin{figure}
  \centering
  \includegraphics[trim=0cm 0cm 0cm 0cm,clip,scale=0.321]{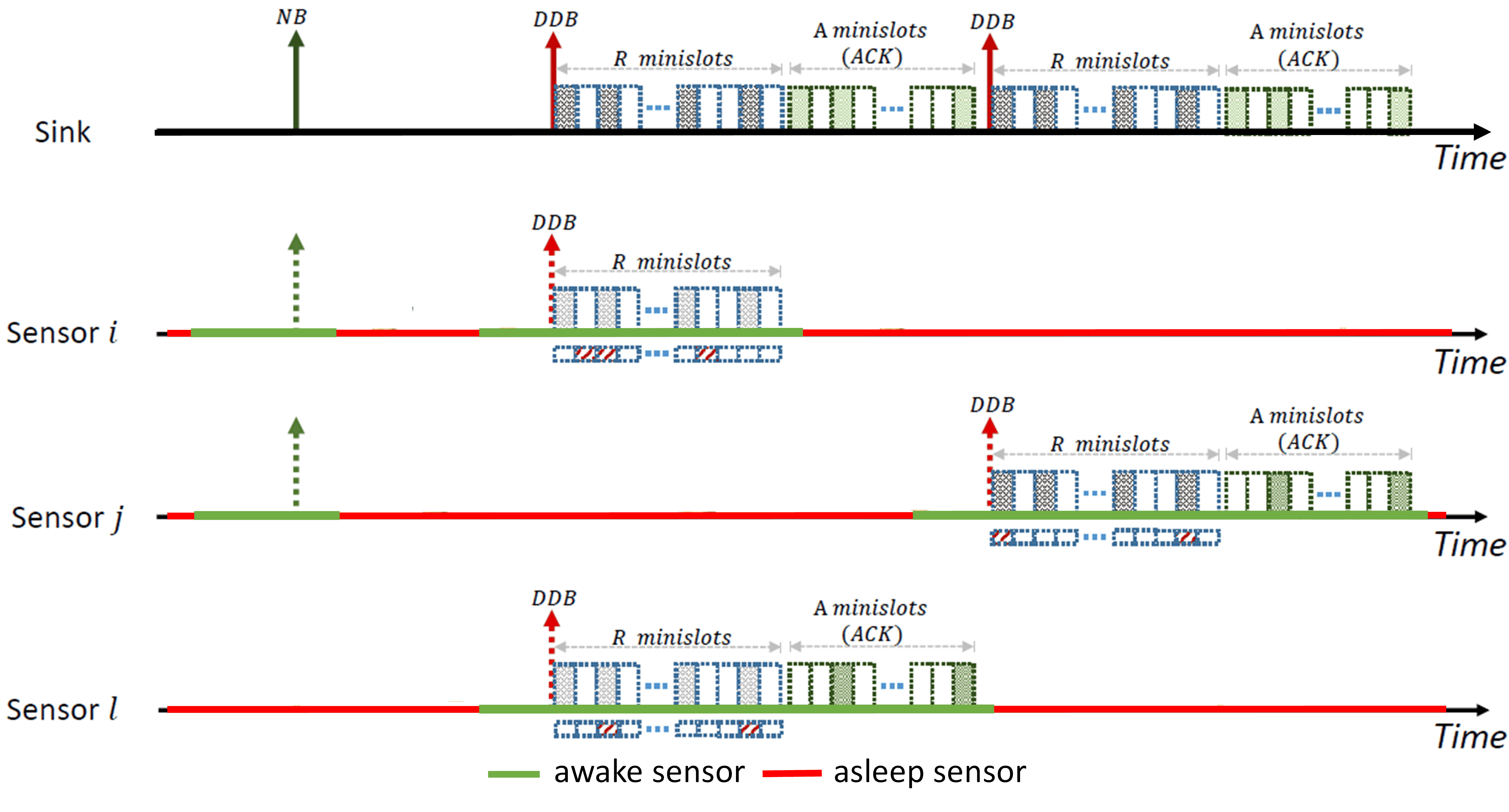}
  \caption{An illustration of downstream protocol operation.}
  \label{fig:DownstreamProtocol}
  \ifshort \vspace{-4mm} \fi
\end{figure}

Each sensor is required to periodically wake up and wait for the sink's beacon. If the beacon is NB, it can go back to sleep. If it is DDB, it stays awake for the $R$ minislot transmission epoch, and after its termination, relying on the observed sequence, it tries to decode whether the collected message conveyed any information to itself (the decoding procedure is presented below). To make sure messages were received, the protocol can rely on ACKs sent by the receiving sensors. These ACKs can be sent simultaneously by all the intended sensors according to the upstream protocol.\off{ This ACK messages are represented by an additional predefined codeword in each sensor's bin of codewords. In any case, since the sink has no indication which of the intended sensors is awake at what time, it is expected to repeat the process for each message until it receives ACK from each intended sensor and up to a limited number of repetitions.} The downstream operation of the protocol is illustrated in Figure~\ref{fig:DownstreamProtocol}.
\subsection{Sink encoding}
Similar to the upstream case, each message is assigned a unique sequence of ones and zeroes of length $R$. The construction of the sequences is random, similar to the one described in Section~\ref{LowerBound}. Accordingly, for each Sensor $j\in \mathcal{K}$, there is a unique codeword $x^{T}(c_j)$ for each of the $c_i$ possible messages. Therefore, the WSN matrix for the predefined downstream messages contains $c_i K$ codewords of length $R$. Assuming the set of intended receivers is $\mathcal{K}$, the sequence transmitted by the sink is
\begin{equation*}
\tilde{x}(l)=\sum_{t=1}^{T}\left(\sum_{j\in \mathcal{K}}X_{j}(t)\right)g(l-(t-1)T_0),\text{ }\text{ } 0 \leq l \leq T \cdot T_0.
\end{equation*}
The channel output signal $\tilde{y}_{j}(l)$ at the $j$-th awake sensor is given by \ifveryshort $\tilde{y}_{j}(l)= h_j \tilde{x}(l) + w_n(l)$. \else
\ifshort
\[
\textstyle \tilde{y}_{j}(l)= h_j \tilde{x}(l) + w_n(l).
\]
\else
\[
\tilde{y}_{j}(l)= h_j \tilde{x}(l) + w_n(l).
\]
\fi
\fi
Hence, the outcome vector $Y(t)\in\{1,\ldots,T\}$ at the awake sensors is $1$ in a minislot $t$ if
\[
\int_{(t-1)T_0}^{tT_0} \tilde{y}_{j}(l)g(l-(t-1)T_0)dl\geq P_{th},
\]
and $0$ otherwise.
Figure \ref{fig:WiretapCoding} depicts the encoding procedure, where in the downstream case, the sink does not  know which of the sensors is awake, yet it transmits the encoded message $\tilde{x}(l)$ assuming that all the addressee sensors are awake.
\subsection{Decoding at the Sensors}
The optimal ML decoder described in Section~\ref{LowerBound} can also be utilized by the sensors for the downstream traffic. However, in order to exploit this ML decoder, each sensor is required to hold the whole code-book, which may require a significantly large memory. Furthermore, the decoding computational complexity can be quite high. These two resources are scarce in WSN.

An alternative decoder is the CoMa decoder described in Section~\ref{efficient_algorithms}. For this decoder, not only the decoding algorithm is computationally efficient, but also each sensor is only required to store a small code-book containing the possible messages destined to itself (obviously, the sink is still required to know all the messages to all sensors). In turn, when receiving the sequence of $R$ mini-slots, each sensor independently tries to match the sequence of received $R$ mini-slots, to one of its own codewords as proposed by the CoMa algorithm; if a codeword matches, the sensor can identify the message associated with this codeword, and can reply with a corresponding ACK. Otherwise, the sensor knows that there was no message destined to itself in this batch.

It is interesting to note that for this downstream protocol, if each sensor utilizes the CoMa decoder, and the sink share with each sensor has only the bin of codewords that are relevant it, one gets privacy, i.e., each sensor can decode only messages destined to itself and cannot decode messages transmitted by the sink to any of the other $K-1$ sensors. The analysis for this case is quite similar to the one provided in \ifsup Appendix \ref{StrongLowerBound}\else Section~\ref{StrongLowerBound}\fi, and the technical enhancements are beyond the scope of this paper.
\ifsup\else
\section{Secure-WSN protocol} \label{StrongLowerBound}
In a Secure-\emph{WSN} protocol, in addition to the goals presented in \Cref{formulation} for the non-secure model, we wish to keep an \emph{eavesdropper}, which may be able to observe a subset of the transmissions, ignorant regarding  \textit{the information transmitted and the information of which subset of $\mathcal{K}$ sensors were active}.
\off{Figure \ref{figure:secure-group-testing} gives a graphical representation of the secure model.
\begin{figure}
  \centering
  \includegraphics[trim= 0cm 0cm 0cm 0cm,clip,scale=0.82]{AdHocNetwork_4.png}
  \caption{Secure Wireless sensors network.}
  \label{figure:secure-group-testing}
  \ifshort \vspace{-3mm} \fi
\end{figure}
}

We assume eavesdropper can observe a noisy vector $\tilde{z}(l)$, generated from the output signal $\tilde{y}(l)$.
In this paper we consider an erasure channel at the eavesdropper, with erasure probability of $1-\delta$, i.i.d. That is, on average, $T\delta$ mini-slots of the channel output are not erased and are available to the eavesdropper.
We assume the eavesdropper uses the same hard decision mechanism, hence observes $Z^T \in \{0,1,?\}^T$. While this is an un-necessary restriction on Eve, it simplifies the \emph{technical} aspects of this section and allows us to focus on the key methods.

The reliability constrain given in \Cref{def:reliable_secure} and the following secrecy constrain lay out the goals of the algorithm.
\begin{definition}\label{def:reliable_secure1}
A sequence of algorithms with parameters $N,K,C$ and $T$ is asymptotically \emph{secure} if, at the eavesdropper, observing $Z^T$, we have
\begin{equation*}
\lim_{T \to \infty}\frac{1}{T}I(\textbf{M}_W;Z^T) = 0.
\end{equation*}
That is, Eve cannot decode anything from the set of messages.
\end{definition}
Thus, observing $Y^T$, in the data collection setup the sink will identify the subset of active sensors and the messages transmitted by them, in the data dissemination setup, each sensor will identify his corresponded messages transmitted by the sink, yet, observing $Z^ T$ in either of the setups, the eavesdropper will not be able to identify the subset of active sensors and the messages transmitted.

The transmission and detection processes are similar to the ones described in \Cref{formulation} and \Cref{Dissemination} for the non-secure protocols, yet, in order to support a secured version of the protocol, we modify the code and the decoding algorithm.

For simplicity, we only describe the secure data collection setup. However, extending the transmission and detection processes described in \Cref{Dissemination} to the secure data dissemination setup is straightforward, since, the code and the decoding algorithm are the same as in the secure data collection algorithm we now describe.
\subsection{Sensors Encoding Process}
In order to keep the eavesdropper ignorant, we propose another level of binning. Specifically, for each sensor $n\in N$, we create a sub-bin for each message.
We then \emph{randomly} map each message that the sensor wants to transmit, $c\in \{1,\ldots,C\}$, to a codeword in the corresponding sub-bin, that is, the $\{n,c\}$-th sub-bin, which contains $F$ codewords of size $T$.
Figure \ref{fig:StrognWiretapCoding} depicts an example.

More precisely, we assume a source of randomness $(\mathcal{R},p_R)$, with known alphabet $\mathcal{R}$ and known statistics $p_R$ is available to the encoder.
It is important to note that this source of randomness does not have to be shared with any other party.
A \emph{stochastic encoder} \cite{C13}, at each active sensor $j\in \mathcal{K}$, selects uniformly at random one codeword $x^{T}(c_j,f)$ , $1 \leq c\leq C$ and $1 \leq f\leq F$ from his sub-bin, that is, it maps a selected message and the source of randomness to a transmission codeword $\textbf{X}^{T}_{n}$.
This mapping, using the randomness, is intended to confuse the eavesdropper regarding the sensor transmitting and the message sent.
Hence, over the MAC channel, we still have a transmission matrix $\textbf{X}^{T}_{S_{w}}$,
each of its rows corresponding to a different active sensor in the index set $S_w$, and that transmission matrix contains $K$ codewords of size $T$,
yet now there is no $1:1$ mapping between messages and codewords, and each message corresponds to $F$ codewords.
\subsection{Codebook Generation}
For each sensor we generate a bin, containing several sub-bins.
The number of such sub-bins \emph{corresponds} to the number $C$ of messages that each sensor has.
The number $F$ of codewords in each sub-bin corresponds to $T\delta$, the number of un-erased mini-slots that the eavesdropper may obtain, yet normalized by the number of active sensors.
The codebook is depicted in the central side of Figure \ref{fig:StrognWiretapCoding}.

Thus, for each sensor we generate a bin of $C \cdot F$ random codewords (the codewords are generated similar to the non-secure model).
Then, we split each bin to sub-bins of $F$ codewords $x^{T}(c,f)$, $1 \leq c\leq C$ and $1 \leq f\leq F$.
Hence, for each message, $c\in\{1,\ldots,C\}$, there are $F$ possible codewords correspond in the $\{n,c\}$-th sub-bin. During the encoding process, only one codeword from the $\{n,c\}$-th sub-bin will be randomly selected for transmission. \textit{We assume Eve may have this codebook as well}.
\subsection{Decoding at the Sink}
The decoder in the SWSN algorithm is similar to the one proposed in \Cref{LowerBoundDec} for the non-secure WSN.
Hence, after $Y^T$ is obtained, the ML decoder looks for a collection of $K$ codewords $\hat{\textbf{X}}_{S_{\hat{w}}}^{T}$,
\textit{each one taken from a separate sub-bin under different bins}, for which $Y^T$ is most likely. Namely,
\ifshort
\begin{equation*}
\textstyle P(Y^{T}|\hat{\textbf{X}}_{S_{\hat{w}}}^{T})>P(Y^{T}|\textbf{X}_{S_{\grave{w}}}^{T}), \forall \grave{w} \neq \hat{w}.
\end{equation*}
\else
\begin{equation*}
P(Y^{T}|\hat{\textbf{X}}_{S_{\hat{w}}}^{T})>P(Y^{T}|\textbf{X}_{S_{\grave{w}}}^{T}), \forall \grave{w} \neq \hat{w}.
\end{equation*}
\fi
However, due the randomness in the encoding procedure of the SWSN algorithm, there is no $1 : 1$ mapping between messages transmitted and codewords, and each message corresponds to a few codewords.
Hence, the sink looks for both the set $\hat{w}$, and the codewords $\hat{\textbf{X}}_{j}^{T}, \forall j\in S_{\hat{w}}$, which are most likely.
Then, the sink declares $\hat{W}(Y^T)$ as the set of active sensor, where $\hat{w}$ is the set of bins in which the codewords reside and maps the selected codewords $\hat{\textbf{X}}^{T}_{S_{\hat{w}}}$ back to the messages $\hat{c}_{j}$ according to the corresponding sub-bins.
\begin{figure}
  \includegraphics[trim= 0.8cm 0cm 0cm 0cm,clip,scale=0.93]{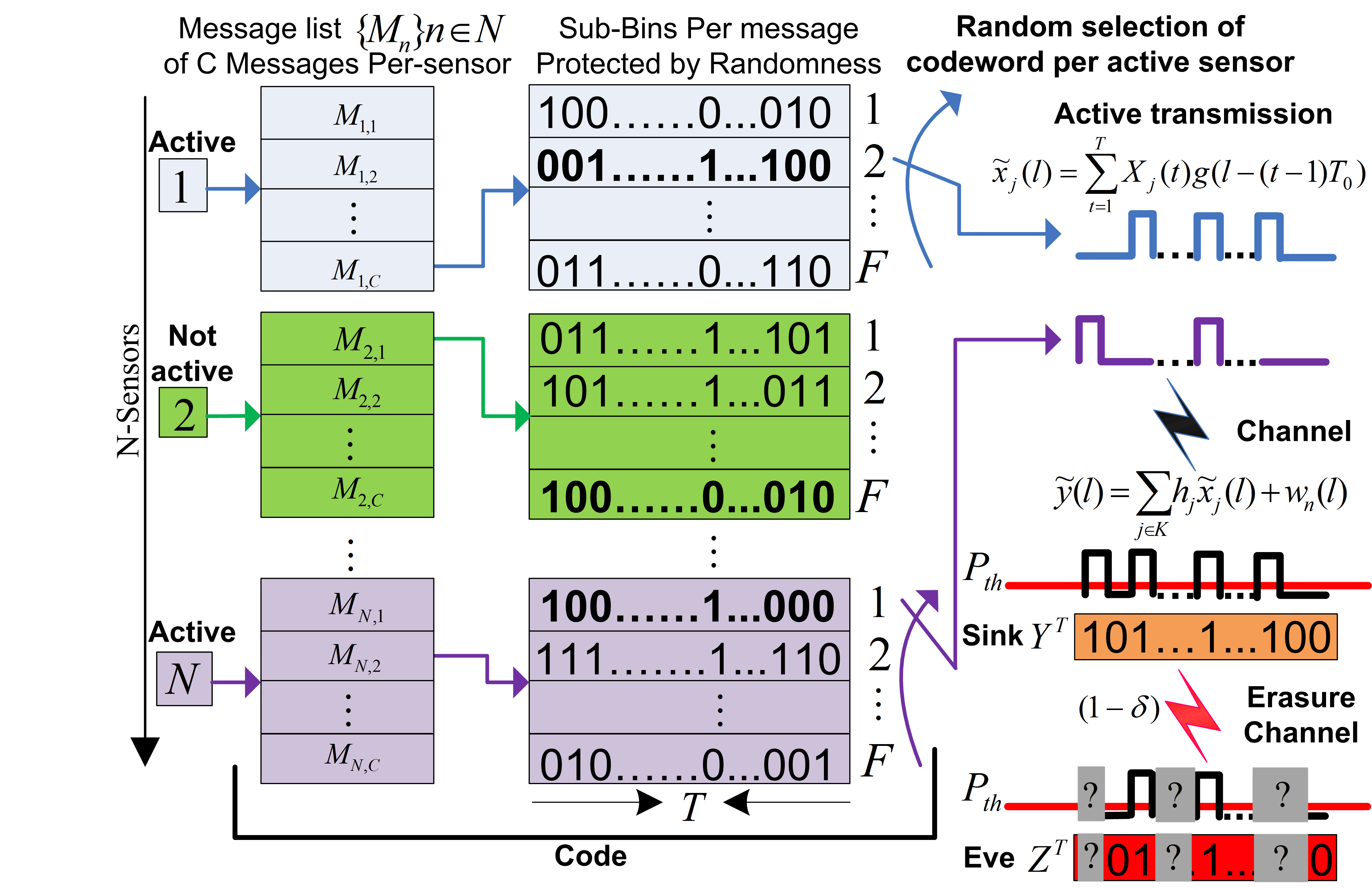}
  \caption{Encoding, transmission and detection in the SWSN protocol.}
  \label{fig:StrognWiretapCoding}
  \ifshort \vspace{-3mm} \fi
\end{figure}
\subsection{Reliability}
The reliability proof using SWSN protocol is almost a direct consequence of the proof given in Section \ref{LowerBound} for the non-secure WSN.
However, there is a main difference: the decoder has $\binom{N}{K}C^KF^K$ possible subsets of codewords to choose from,
$\binom{N}{K}$ for the number of possible sensors, $C^K$ for the number of possible messages in each bin and $F^K$ for the number of possible codewords to take in each sub-bin.
Thus, when fixing the error event, there are $\binom{N-K}{i}C^KF^K$ subsets to confuse the decoder.
Specifically, to obtain the bound on the size of the codewords given in Lemma \ref{direct lemma2}, the error probability analysis in the ML decoder extends the bound given in \ifveryshort \cite[Lemma 3]{wsn2017drivejornal}\else \Cref{error lemma2 with E'}\fi, by considering $C \cdot F$ codewords per sensor,
and by considering multiple error events, as given in \cite{sgt2016arxiv}, for the analysis of the secure GT error probability bound.
E.g., events where the decoder chooses the wrong codeword for some sub-bin, yet identified the sensors and the messages transmitted correctly (since the sub-bins were correctly identified),
and events where the codeword selected was from a wrong sensor sub-bin (hence resulted in an error).
\begin{lemma}\label{direct lemma2}
If the size of the codewords satisfies
\begin{eqnarray}\label{eq:reduce_h}
T & \ge & \max_{1 \leq i \leq K} \frac{1}{1-(1+\varepsilon)\delta}\frac{1+\varepsilon}{i/K}\log\binom{N-K}{i}C^i,
\end{eqnarray}
then, under the codebook above, as $N\rightarrow \infty$ the average error probability approaches zero.
\end{lemma}

The analysis of the information leakage at the eavesdropper, to prove the security constraint is met,
is a direct consequence of the proof given in \cite[Section V.B]{cohen2016secure}, where we protect not only on the information of which of the sensors was transmitting a message
but also protect the messages as well. Hence, in the same way as given in \cite{cohen2016secure}, yet, instead of showing that $I(W;Z^{T})/T\rightarrow 0$,
we can show that $I(\textbf{M}_W;Z^{T})/T\rightarrow 0$.

To conclude, assuming a SWSN algorithm with the parameters $N$, $K$ and $C$, for any $0\leq \delta < 1$, if size of the codewords
\ifshort
\begin{equation}\label{Sorder_T}
\textstyle   T = \Theta \left(\frac{K \log N C}{1-\delta}\right),
\end{equation}
\else
\begin{equation}\label{Sorder_T}
   T = \Theta \left(\frac{K \log N C}{1-\delta}\right),
\end{equation}
\fi
for some $\varepsilon \geq 0$, then there exists a sequence of SWSN algorithms which are reliable and secure (under the conditions given in Definition \ref{def:reliable_secure} and Definition \ref{def:reliable_secure1}). Again, using the upper bound $\log\binom{N-K}{i} \leq i\log \frac{(N-K)e}{i}$, now on \Cref{direct lemma2}, the maximum over $i$ is easily solved, and we have
\[
T \geq  \frac{1+\varepsilon}{1-\delta} K \log (N-K)Ce.
\]             %
\fi
\section{Advance Techniques For Wireless Transmission} \label{Practical}
To illustrate the WSN protocol suggested herein for wireless radio transition, we assumed that sensors (in the data collection setup) or the sink (in the data dissemination setup), utilize a simple PAM. However, it is important to note that the protocol suggested can utilize different transmission techniques. In this section we will explore the utilization of the Orthogonal Frequency Division Multiple Access (OFDMA) \cite{cox2012introduction}. Recall that traditional OFDMA, allows simultaneous low-data-rate transmission to and from multiple users, by allocating different user different fractions of the bandwidth.  We assume that in total there are $F_{ch}$ such closely spaced orthogonal sub-carriers which are used to carry the data, and denote each by $f_{ch}\in \{1,\ldots,F_{ch}\}$ .

\begin{figure}
  \centering
  \includegraphics[trim= 0.1cm 0cm 0cm 0cm,clip,scale=0.8]{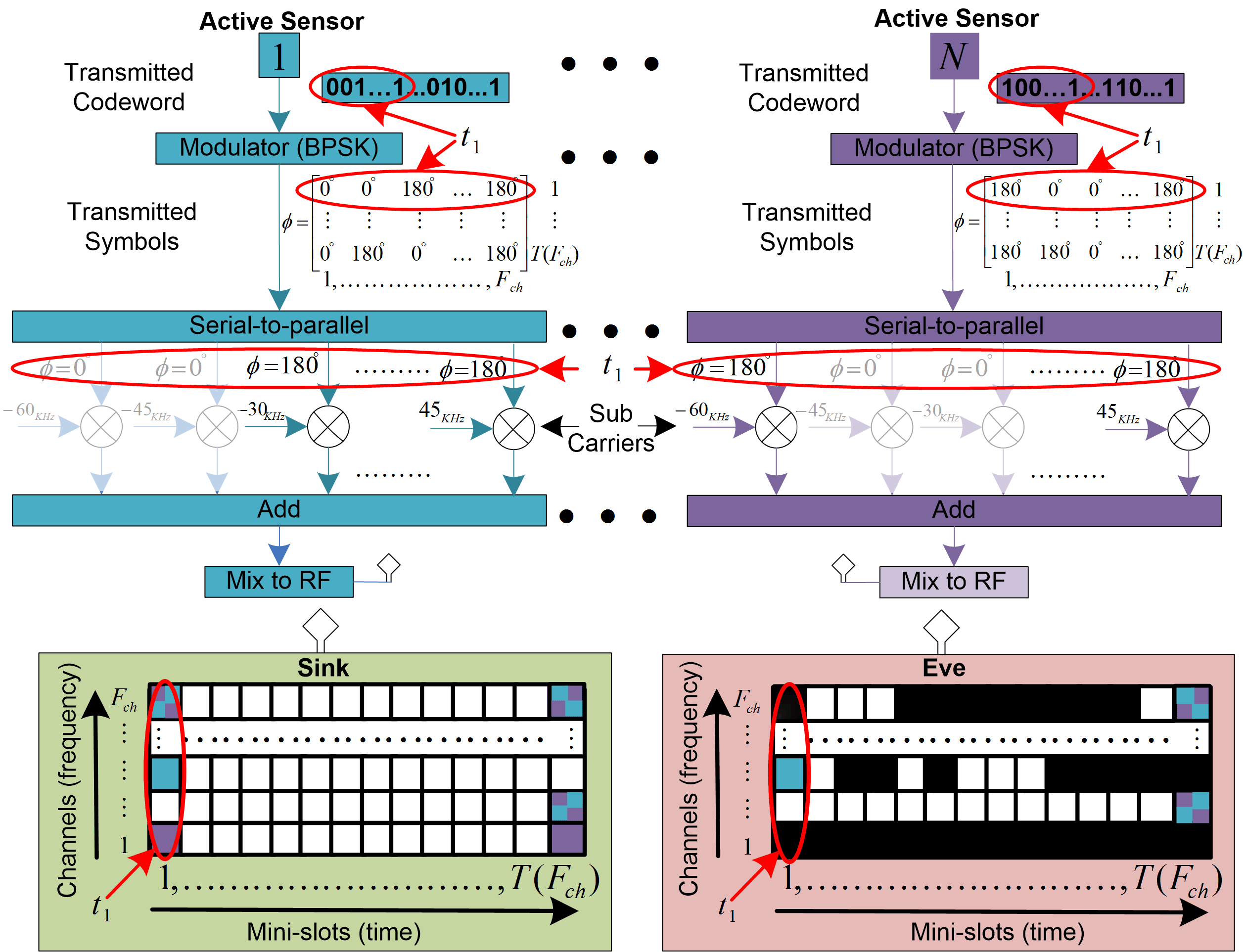}
  \caption{Transmission using a modified OFDMA technique for the suggested protocols.}
  \label{fig:OFDMA}
  \ifshort \vspace{-3mm} \fi
\end{figure}

In order to support the WSN protocol suggested in \Cref{sec:model,LowerBound}, we devised a modified OFDMA technique, in which there is no \emph{a-priori} sub-carrier assignment to the active sensors, as in traditional OFDMA, but instead each of the $K$ active sensors can transmit its codeword on all the sub-carriers. Specifically, each active sensor maps its codeword to the subcarriers. In particular, the sensor transmits on the sub-carriers $f_{ch}$ that correspond to indices which are 1's in the codeword it intends to transmit, and does not transmit on the sub-carriers that correspond to indices which are 0. For example, if the codeword the sensor intends to transmit is $001001$ it will transmit in the third and sixth sub-carriers and will not transmit in the others. After mapping the codeword to the sub-carriers, the sensor follows the regular FDMA encoding procedure \cite{cox2012introduction}. A graphical illustration of the modified OFDMA using BPSK modulator at each active sensor is given in \Cref{fig:OFDMA}. Obviously, if the codeword is greater than the number of sub-carriers ($T>F_{ch}$) it needs to be partitioned and transmitted in $T/F_{ch}$ minislots. Accordingly, assuming a WSN algorithm given in \Cref{LowerBound}, with the parameters $N$, $K$ and $C$, and using OFDMA and BPSK techniques, with $F_{ch}$ frequency channels, returning to \eqref{order_T} the number of mini-slots required to transmit the codewords is
\ifsup
$T(F_{ch})=\Theta \left(\frac{K \log N C}{F_{ch}}\right)$.
\else
\[
 T(F_{ch})=\Theta \left(\frac{K \log N C}{F_{ch}}\right).
\]
\fi
Obviously, the same procedure also applies to the Data Dissemination protocol (\Cref{Dissemination}). Only this time the sink maps each of the codewords it intends to transmit into the sub-carriers and performs the OFDM procedure over the collected mapping, in which if a sub-carrier is mapped by two or more codewords, the sink refers to it just as if it was mapped by a single sub-carrier.

It is important to note that the modified OFDMA suggested herein also applies to the Secured-WSN version suggested in \ifsup Appendix \ref{StrongLowerBound}\else\Cref{StrongLowerBound}\fi. Under this secured version we assume that due to interference, collisions (low SINR) or low SNR, the eavesdropper is not accessible to all the frequency channels in each mini-slot. Specifically, the eavesdropper is available to at most $F_{ch}T(F_{ch})\delta$ symbols transmitted. Hence, assuming a SWSN algorithm suggested in \ifsup Appendix \ref{StrongLowerBound}\else\Cref{StrongLowerBound}\fi, with the parameters $N$, $K$ and $C$, for any $0\leq \delta < 1$, and using OFDMA and BPSK techniques, with $F_{ch}$ frequency channels, according to \eqref{Sorder_T} the number of mini-slots required to transmit the codewords is
\ifsup
$T(F_{ch})=\Theta \left(\frac{K \log N C}{F_{ch}(1-\delta)}\right)$.
\else
\[
 T(F_{ch})=\Theta \left(\frac{K \log N C}{F_{ch}(1-\delta)}\right).
\]
\fi
       %
\section{Simulation Results} \label{Sim}
In this section, we provide insight into the proposed protocol and the analytical results presented in this paper, both for the unsecured and secured versions, via simulations. It is important to note that comparison with all-purpose WSN MAC protocol is pointless, as the overhead required to support such all-purpose protocols is one order of magnitude greater than the one required in our designated protocol. For example, utilizing RI-MAC \cite{sun2008ri}, which is considered a highly efficient protocol, for the data harvesting problem presented in this paper, the sink node would need to broadcast a beacon, similar to the RFR herein, followed by the report transmissions by all pending sensors. Obviously, if more than one sensor is waiting for transmission, which is highly probable in such a dense network, a collision will occur and a tedious collision resolution will follow; in case of several nodes transmitting simultaneously, this  can repeat several times, which is time consuming. Besides the overhead and latency consumed due to the channel access and collision resolution mechanism, the message payload, which can range between several bytes per message to much longer messages, is also airtime consuming (e.g., in \cite{sun2008ri} the performance on short packets of $28 Bytes$ was examined). In addition, one needs to take into account the overhead induced by data encapsulation which can include multiple addresses and other control information (e.g., an $802.11$ ACK message, which conveys only receiver ID and a single bit of information is $14 Bytes$ plus physical-layer encapsulation). In contrast to all this overhead, as we will show in our results, assuming a network of $500$ nodes and allowing up to $5$ sensors transmitting simultaneously, each having up to $10$ different messages it can transmit, will require only about $65$ minislots, each of which can be the length of a \emph{single modulated bit.}

All simulations presented are conducted in MATLAB, where we implemented the suggested protocols in full: code construction, encoding, the multiple access wireless channel (noise and propagation channel gain) and decoding procedure at the sink, as well in the secure model decoding procedure at the eavesdropper.

First, we examine the noiseless channel assuming that there are no errors in the signal received. In \ifshort Figures~\ref{fig:DND_simulation}(a) and~\ref{fig:DND_simulation}(b) \fi we examined the size of the required codeword ($T$) to decode the messages, for the unsecured and secured versions, respectively. In particular, for $N=500$, $C=10$ and with $K=3$, we repeated the simulations trying to decode the messages for different $T$ and presenting the minimal $T$ required both for the ML and the CoMa decoders. It is important to note that in the secure model we adapted the number of codewords in each sub-bin as required in the code construction, to keep the eavesdropper completely ignorant, namely, with success probability approaching zero, i.e., for each $T$ we constructed a different code. Each plotted point in the figure depicts the average of $4000$ simulations.
\ifshort
\begin{figure}
  \centering
  \includegraphics[trim= 6.2cm 1.4cm 7.2cm 0.5cm,clip,scale=0.250]{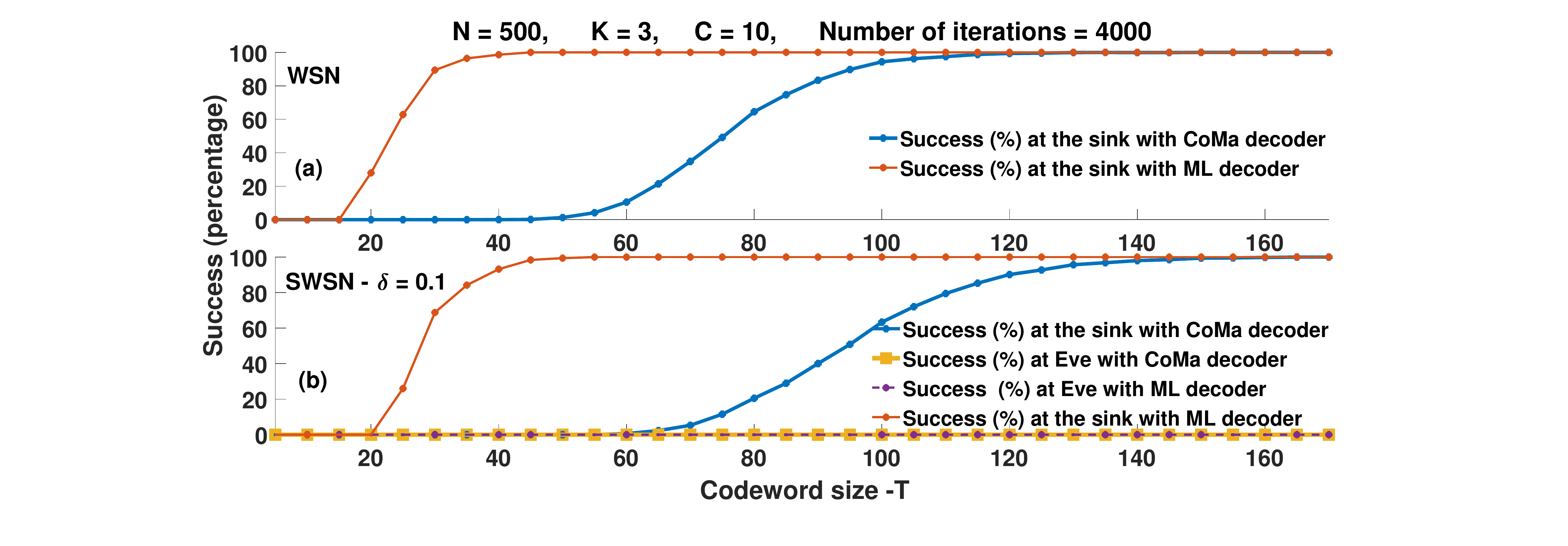}
  \caption{ML and CoMa simulation results. (a) WSN protocol (b) SWSN protocol.}
  \label{fig:DND_simulation}
  \vspace{-3mm}
\end{figure}
\fi

Figure~\ifshort \ref{fig:DND_simulation}(a) \fi depicts the performance of the WSN model given in Sections \ref{LowerBound} and \ref{efficient_algorithms}, using CoMa and ML decoding algorithms. In particular, we present the Cumulative Distribution Function (CDF) of decoding the messages as a function of the size of the codeword in the code ($T$). For example, as can be seen in the figure, in order to attain $95\%$ success probability the required codeword length is $T = 35$ and $T = 105$ for the ML and the CoMa decoders, respectively, i.e., in $95\%$ of the simulation instances we succeeded to decode the messages for these $T$'s. In order to attain $100\%$ of success, $T$ should be $45$ for ML and $130$ for the CoMa decoder. It is important to note that for this setup there are no false negatives, i.e., the sink can mistakenly decode more messages than actually sent, but will not miss a message. Accordingly, one can compromise the success probability slightly in order to shorten the transmission time.

In Figure~\ifshort \ref{fig:DND_simulation} (b) \fi we show the performance of SWSN model given in \ifsup Appendix \ref{StrongLowerBound}\else Section \ref{StrongLowerBound}\fi, using the same setup as in the previous simulations, only this time in the presence of  an eavesdropper which is accessible on average to $T\delta$ mini-slots from the sink output channel, where $\delta=0.1$ (i.e., in the simulations we have randomly decided whether the eavesdropper can sense the energy generated on each minislot, such that each minislot was heard by the eavesdropper with probability $\delta=0.1$, independent from one minislot to the other). Again, it can be seen that the simulation results agree with the analytical results given in \ifsup Appendix \ref{StrongLowerBound}\else Section~\ref{StrongLowerBound}\fi. Specifically, the code length, $T$, (number of minislots) which guarantees $100\%$ decodable rate is $T=55$ and $T=160$ for the ML and the CoMa decoders, respectively, falling behind the analytical upper bound.

Next, we provide numerical results to provide insight into the tradeoffs between the different parameters, and specifically the number of sensors, the number of active sensors and the sequence length. Note that the following results only present the analytical upper bound of Section \ref{LowerBound}, which as can be seen, are typically pessimistic.

Figure \ifshort\ref{fig:NK_SNK_simulation}, left panel, \fi depicts the relation between the size of the codewords $T$ and the number of sensors $N$ for different $K$s, in the unsecured version. Namely, for each line in the figure, we keep $K$ and $C$ fixed, and vary $N$ to show the bound on the codeword length, $T$, according to \eqref{eq:reduce_hw}. It can be clearly seen that the rate in the network scales logarithmically with $N$ for all considered $K$, in contrast with TDMA, where the number of slots required scales linearly with the number of transmitting sensors. Accordingly, we can conclude that as long as the number of simultaneously transmitting nodes and the number of different messages each sensor stores are kept fixed, the suggested protocol scales very well with the number of sensors in the network, i.e., the protocol is efficient even when there is a high number of sensors in the network.
\ifshort
\begin{figure}
  \centering
  \includegraphics[trim= 5.2cm 0cm 5.0cm 1.0cm,clip,scale=0.230]{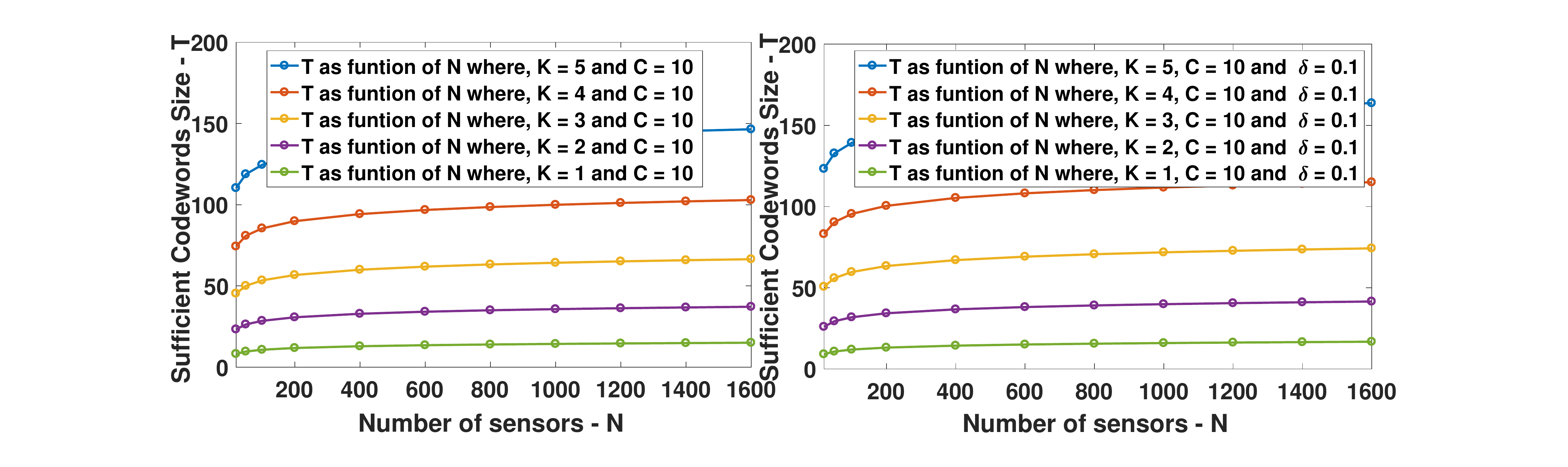}
  \caption{Numerical results of the achievable schemes, for different values of N and K. Left panel WSN, right panel SWSN.}
  \label{fig:NK_SNK_simulation}
  \vspace{-4mm}
\end{figure}
\fi

Figure \ifshort\ref{fig:NK_SNK_simulation}, right panel, \fi according to \eqref{eq:reduce_h}, for the
depicts the same setup for the secured version (SWSN), for $\delta =0.1$, i.e., the eavesdropper is accessible to $10\%$ of the mini-slots. Note that even though the rate required in order to keep the eavesdropper ignorant is slightly higher with respect to the unsecured version, the same logarithmic trend in $N$ is kept.

\ifshort Figures~\ref{fig:KN_SKN_simulation} \fi depicts the results when we keep $N$ and $C$ fixed and vary the number of simultaneously transmitting sensors, $K$, for different $N$s. The achievable rate in the network is linear in $K$, hence can be quite effective even for larger $K$. Recall that $K$, the number of simultaneous transmissions, affects the latency of the reports. E.g., as the number of simultaneous attempts rises, a protocol with exponential random back-offs will suffer significantly from both bad channel utilization and as high latencies. In the suggested protocol, the degradation is graceful.
\ifshort
\begin{figure}
  \centering
  \includegraphics[trim= 5.1cm 0cm 5.0cm 1.0cm,clip,scale=0.230]{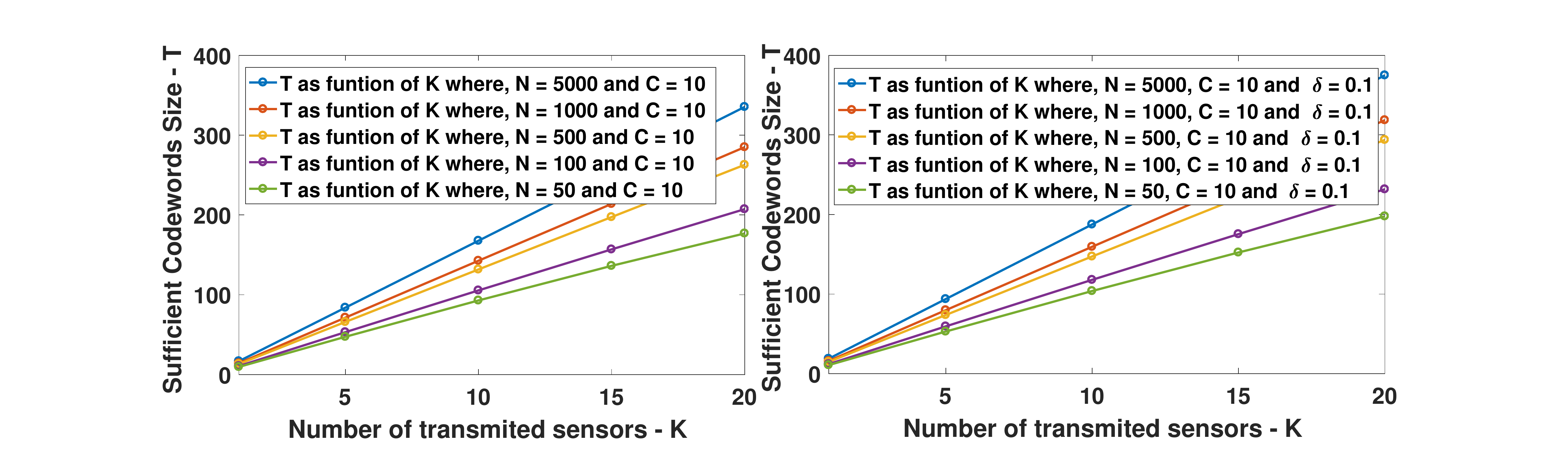}
  \caption{Numerical results of the achievable schemes, for different values of $N$ and $K$. Left panel WSN, right panel SWSN.}
  \label{fig:KN_SKN_simulation}
  \vspace{-4mm}
\end{figure}
\begin{figure}
  \centering
  \vspace{-4mm}
  \includegraphics[trim= 5.2cm 0cm 5.0cm 1.0cm,clip,scale=0.230]{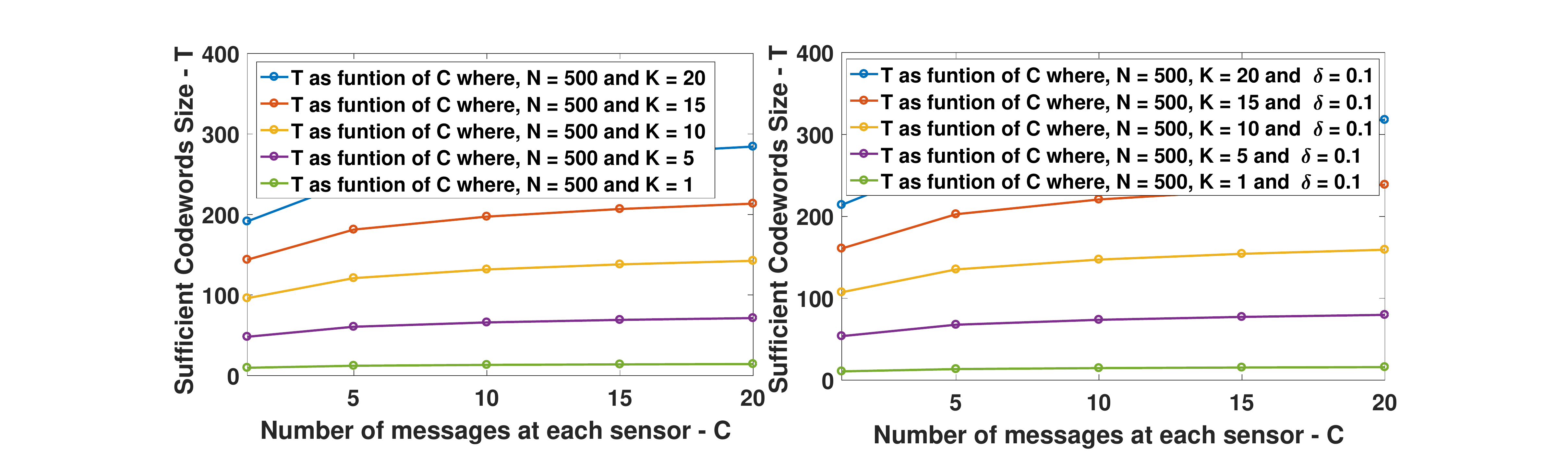}
  \caption{Numerical results of the achievable schemes, for different values of $C$ and $K$. Left panel WSN, right panel SWSN.}
  \label{fig:C_SC_simulation}
  \vspace{-3mm}
\end{figure}
\fi

\ifshort Figures~\ref{fig:C_SC_simulation} \fi depict the results when $N$ and $K$ are fixed and the number of messages each sensor has ($C$) is varied, for various $K$s. The figure clearly depicts that the logarithmic trend in the achievable rate as a function of $C$, is kept both for the unsecured and secured versions, meaning that when the number of possible messages each sensor can send increases, the suggested protocol is still efficient. \ifveryshort\else However, one needs to recall that the number of possible messages grows exponentially with the message size, i.e., each bit in the payload multiplies the message bank by two. Accordingly, trying to adopt the procedure to the traditional messages is definitely not scalable.\fi

To conclude, the numerical results illustrate trends in the rates of the protocol, which are consistent with the analytical results presented in Section \ref{LowerBound} and Appendix \ref{StrongLowerBound}. We note that the protocol is effective even when the number of sensors in the network or the number of simultaneous transmitting sensors is high, as long as the number of messages each sensor can transmit is moderate. \rev{In Appendix \ref{Sim_app}, we present simulation results for the complete secure protocol. Moreover, in \Cref{sec:Imp}, we present an implemented wireless setup with GNU Radio and RF hardware, in which we validate and obtain the numerical results given in this section.}
\vspace{-0.2cm}
\section{\revised{Implementation}}\label{sec:Imp}
\rev{In this section, we present two implementations under a complete setup of the suggested protocol and the efficient decoding algorithm based on Sections IV-VII. In the first implementation, we validate the suggested wireless protocol, utilizing GNU Radio open-source software development toolkit (\hspace{-0.01mm}\cite{GnuRadio}) with external RF hardware. In the second implementation, we utilize low cost devices to further demonstrate that the suggested protocol can operate efficiently in real time on low cost devices with highly limited capabilities (power, memory and computation limitations).}

\begin{figure}
  \centering
  \includegraphics[trim= 0.5cm 0.0cm 0cm 0cm,clip,scale=0.3]{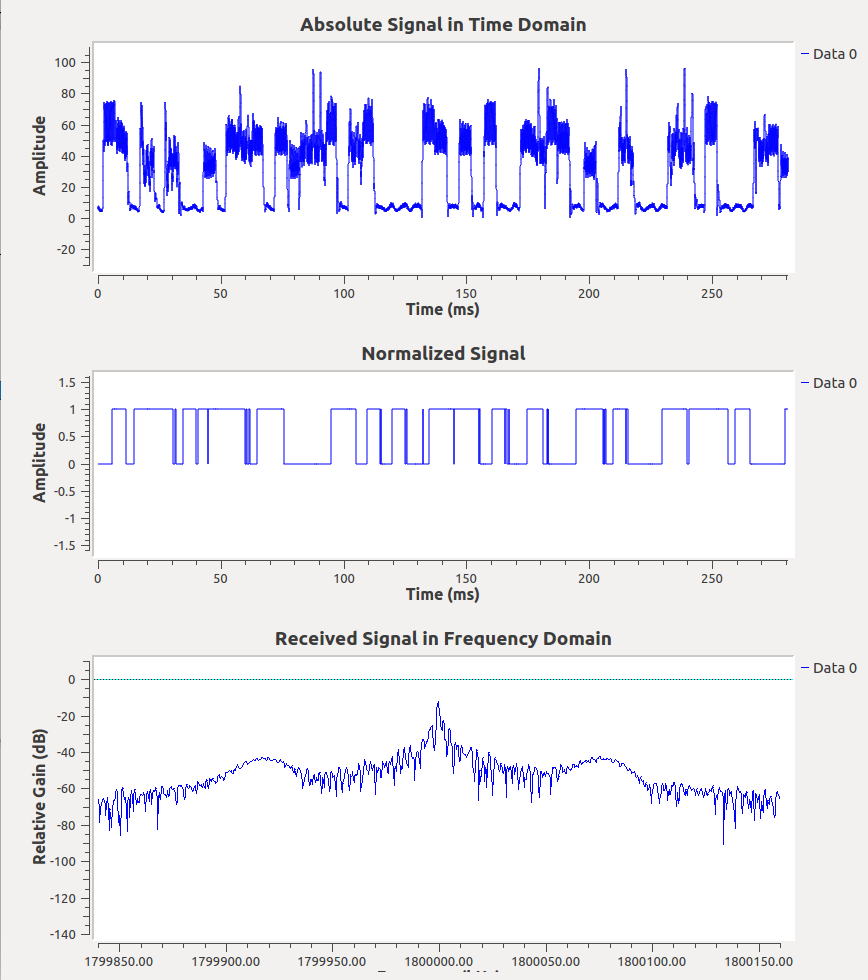}
  \caption{Representation of the received signal. (a) Upper graph, the received signal in time domain. (b) Middle graph, the normalized representation of the received signal after hard decision mechanism. (c) Lower graph, the received signal in frequency domain.}
  \label{figure:gnu_received_signal}
  \vspace{-4mm}
\end{figure}

\rev{In order to demonstrate the suggested protocol’s operation over the air, we have implemented both the sink and sensors on the USRP Software Defined Radio Devices $NI-USRP-2901$.\off{ We have implemented the complete protocol as described in Sections IV-VII.} In our implementation each sensor held a bank of $10$ different messages ($C=10$) and allowed $3$ randomly selected devices to transmit simultaneously ($K=3$). The code was designed for $50$ devices and the transmission interval was $65$ minislots ($T=65$). The signals were transmitted over RF frequency of $1.8GHz$ with $100Hz$ bandwidth, utilizing PAM modulation. The sensors’ sampling rate is $160KHz$, whereas the signal received at the sink after Automatic Gain Control (AGC) is of power below $100mW$ and sampled at a rate of $320KHz$. The sink’s sample rate is $1600$ samples per mini-slot. We have implemented the CoMa decoding scheme which despite the over the air transmissions experienced very high accuracy both in detecting the transmitting devices and the correct message transmitted, consistent with the analytical results presented in Section~\ref{LowerBound}. Figure \ref{figure:gnu_received_signal} depicts the received signal from the channel in the time domain, its corresponding square signal after hard decision mechanism and the received signal in frequency domain.}

\begin{figure}
  \centering
  \includegraphics[trim= 0cm 0.5cm 0cm 0cm,clip,scale=0.32]{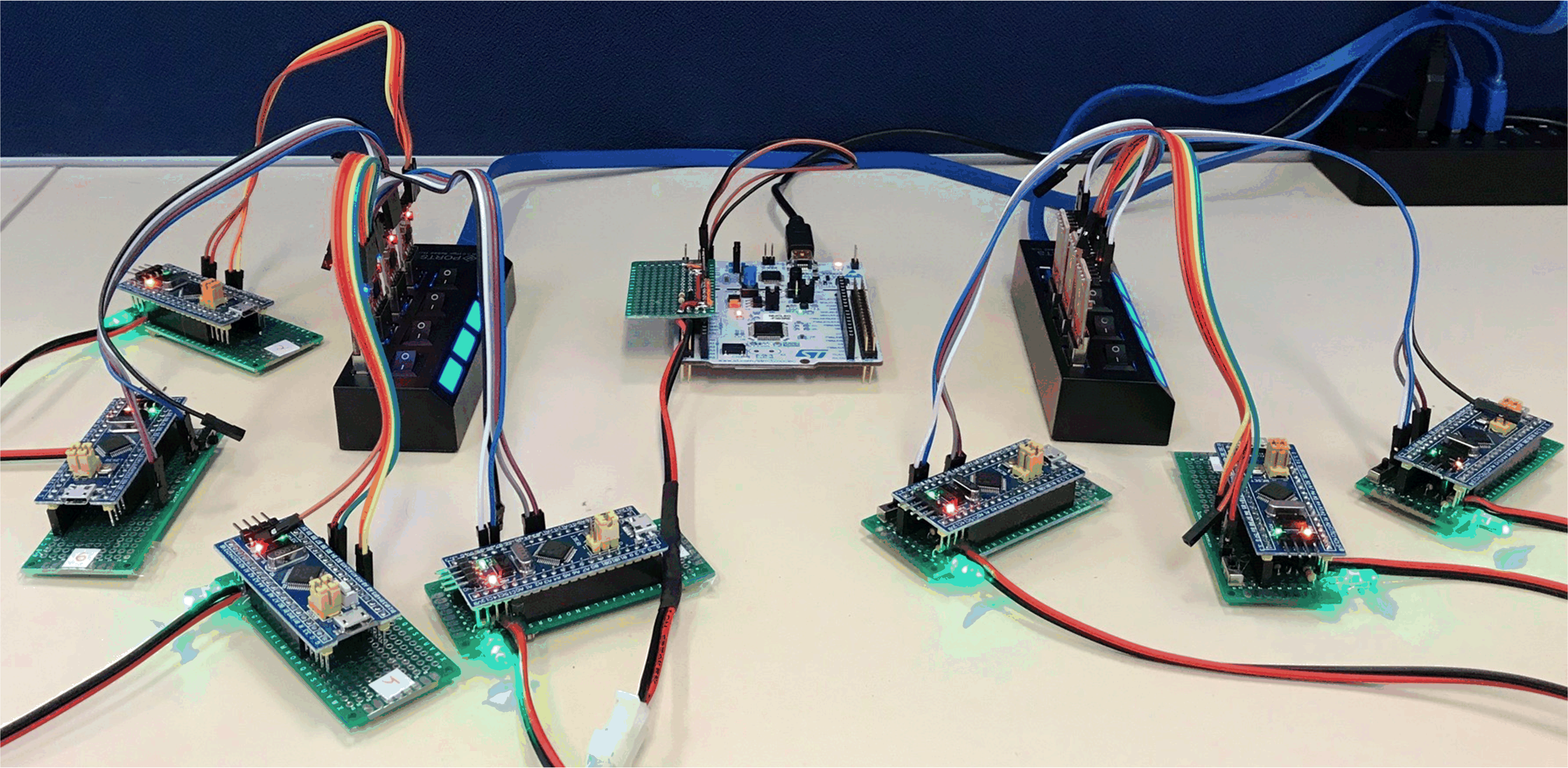}
  \caption{Hardware implementation.}
  \label{figure:hw}
  \vspace{-4mm}
\end{figure}

\rev{In the second implementation} we demonstrate the operation of the protocol on low cost devices. We have implemented a sink and seven sensors using $STM32F303RE$ and $STM32F103C8T6$ evaluation boards, respectively. The sensors and the sink were connected to a joint (wired) bus (Figure \ref{figure:hw}). Each sensor sporadically chose one out of $35$ preloaded messages and waited for transmission. The sink periodically transmitted an RFR signal ($6$ bits at $45 Hz$) and the sensors with pending messages started transmission following the sink’s signal. The codewords were of length $T = 250$, transmitted at $5 Hz$. Note that the slow transmission rate was chosen solely to be able to visualize the transmitted signal using the green LEDs. Once the (combined) signal was received at the sink, it initiated a decoding procedure based on the efficient decoding algorithm given in \Cref{efficient_algorithms}. This implementation demonstrates that even very simple sensors, with a CPU speed of only $8 MHz$ and $8KB$ of memory, were able to carry out the protocol successfully.
\vspace{-0.1cm}
\ifsup\else
\section{\revisedb{Error Probability Bound}}\label{appendix:APPENDIX A}
In this section, we show a sketch of the bound on the error probability of the maximum likelihood decoder based on the proof given in \cite{atia2012boolean,cohen2016secure}.
Thus, under the assumption that all sensors and the messages are equality likely to be active and transmitted, respectively,
we consider the error probability given that the first $K$ sensors are active, that is, $W=S_1$ is the active set. Denote this probability by $P_{e|1}$. We have
\begin{equation*}
P_{e|1} \leq \sum_{i=1}^{K}P(E_i),
\end{equation*}
where $E_i$ is the event of a decoding error in which the decoder declares a wrong message set from the sensors active which differ from the true one in exactly
$i$ transmitted messages from the active sensors ($S_1$).

In the code construction suggested in Section \ref{LowerBound}, for each sensor, there are $C$ possible codewords.
Only one of these codewords is selected by the sensor to be transmitted at each time slot.
Since the decoder does not know which codewords were selected in each bin to be transmitted, there are multiple error events we should consider. E.g.,
events where the decoder chooses the wrong codeword for some sensor, yet identified parts of the sensors correctly, and, events where the codeword selected was from a wrong sensor bin.

In this paper, we consider the error event $E_i$ as the event where a codeword decoded by the sink is not from the set of the codewords transmitted from the active sensors set.
Hence, to avoid error event, in our analysis the decoder, not only got correct codeword in each such bin right, but also the true active sensor is detected by the sink.

Assuming the above definition of the error event, the bound $P(E_i)$ is almost a direct consequence of the proof given in \cite{cohen2016secure},
where in the suggested coding sachem herein, each sensor $n\in N$ has $C$ codewords in his bin.
Particularly, we will establish the following lemma given the proof in \cite[Lemma 3]{cohen2016secure}.
\begin{lemma}\label{error lemma2 with E'}
The error probability $P(E_{i})$ is bounded by
\begin{equation*}
P(E_{i}) \leq 2^{-T\left(E_{o}(\rho)-\rho\frac{\log\binom{N-K}{i}C^i}{T}-\frac{\log\binom{K}{i}}{T}\right)},
\end{equation*}
where the error exponent $E_{o}(\rho)$ is given by
\begin{multline}
E_{o}(\rho)= -\log \sum_{Y\in \{0,1\}}\sum_{X_{\mathcal{S}^2}\in \{0,1\}}\Bigg[\sum_{X_{\mathcal{S}^1}\in \{0,1\}}P(X_{\mathcal{S}^1})\\
p(Y,X_{\mathcal{S}^2}|X_{\mathcal{S}^1})^{\frac{1}{1+\rho}}\Bigg]^{1+\rho},\text{ } 0 \leq\rho \leq1.\nonumber
\end{multline}
\end{lemma}
\fi
\vspace{-0.1cm}
\section{\revisedb{Conclusions}}\label{conc}
In this paper, we design a highly efficient WSN MAC protocol, to collect information from a large number of sensors.
This WSN protocol is specified by a dense set of wireless sensors in the network, out of which a\off{ small} unknown subset may be active and transmit at the same time to a sink.
To support the suggested protocol, we provided a very simple codebook construction with very simple and efficient encoding and decoding procedures.
Moreover, we extend the WSN protocol suggested to a secure version, in which an eavesdropper which has access to a noisy vector from the sink output, will be kept completely ignorant regarding the subset of sensors transmitting and their messages.

To validate the efficiency, of both the non-secure and secure WSN models suggested, we provide rigorous analysis, numerical results and simulations, which illustrate how the different parameters of the network affect the performance of the protocol, namely, the rate of the wireless channel.
\vspace{-0.2cm}                    %
\section*{\revised{Acknowledgments}}
\revised{We would like to thank \rev{Netanel Weiss, Aviad Firuz (R.I.P.),} Tomer Nagar and Shaun Tzirkel for contributing the hardware implementation of the suggested protocol, the related software and testing.}
\vspace{-0.2cm}                       %
\bibliographystyle{IEEE}
\bibliography{references,SecureNetworkCodingGossip}
\ifsup
\newpage
\pagenumbering{arabic}
\setcounter{page}{1}
\begin{center}
  \huge Supplementary Materials
\end{center}
\appendices
             %
\section{Simulation Results - \rev{Secure Protocol}} \label{Sim_app}
To illustrate the all procedure we validate the \rev{secure} protocol under a complete setup and a noisy channel. We examine $N=500$ sensors transmitting to a single sink node; we assume that each sensor holds a bank of $10$ different messages. All wireless channel properties and parameters such as power, distances and noise floors which we used throughout this simulation are typical to such networks according to \cite{cisco2017site}. Sensors used PAM modulation and we have simulated the whole procedure, including the random code generation, the random wakeup procedure, the transmission, the threshold based filtering and the decoding procedure both by the sink and an eavesdropper.

Figure \ref{fig:Simulation_Real_Channel} depicts a random RFR sent by the sink which is followed by a transmission from $2$ sensors (Sensors $177$ and $238$ in this particular instance). Figure~\ref{fig:Simulation_Real_Channel} (a) and (c) depict the two codewords sent by the two sensors (b) and (d) depict the associated PAM modulated signal, respectively. The two sensors were $5m$ and $3m$ from the sink, and obtained power signal at the sink of $17.5dBm$ and $21.1dBm$, respectively (transmission power was $-55dBm$). Figure~\ref{fig:Simulation_Real_Channel}(f) presents the signal $\tilde{y}(l) $ at the sink, which is the channel output resulting from both codewords transmitted simultaneously plus the noise (which was additive white Gaussian noise in $-90dBm$). Note that since there is no need for any gain control and all the more so for automatic gain control (AGC), the received signal strength may vary beyond the receiver saturation point, leading to signal clipping. Figure~\ref{fig:Simulation_Real_Channel}(f) also illustrates the threshold based mechanism, which determines on which of the minislots there was a transmission. The outcome of this hard decision process, which is the input to the decoder (ML or CoMa) is presented in \ref{fig:Simulation_Real_Channel}(e). In this example, the number of minislots utilized is above the upper bound requirement given in \Cref{LowerBound}, thus, the sink can successfully decode both transmitted codewords.

Figure~\ref{fig:Simulation_Real_Channel}(h), presents the signal $\tilde{z}(l)$ received by the eavesdropper, which is accessible on average to $0.25\%$ of the minislots.
The outcome of the hard decision process, which is the input to the decoder (ML or CoMa) at the eavesdropper is presented in Figure \ref{fig:Simulation_Real_Channel}(g).
In the secure model presented in this example, we adapted the number of codewords in each sub-bin as required in the code construction given in \ifsup Appendix \ref{StrongLowerBound}\else\Cref{StrongLowerBound}\fi, to keep the eavesdropper completely ignorant.
\ifveryshort
\begin{figure}
  \centering
  \includegraphics[trim= 6.9cm 1.95cm 19cm 1.6cm,clip,scale=0.370]{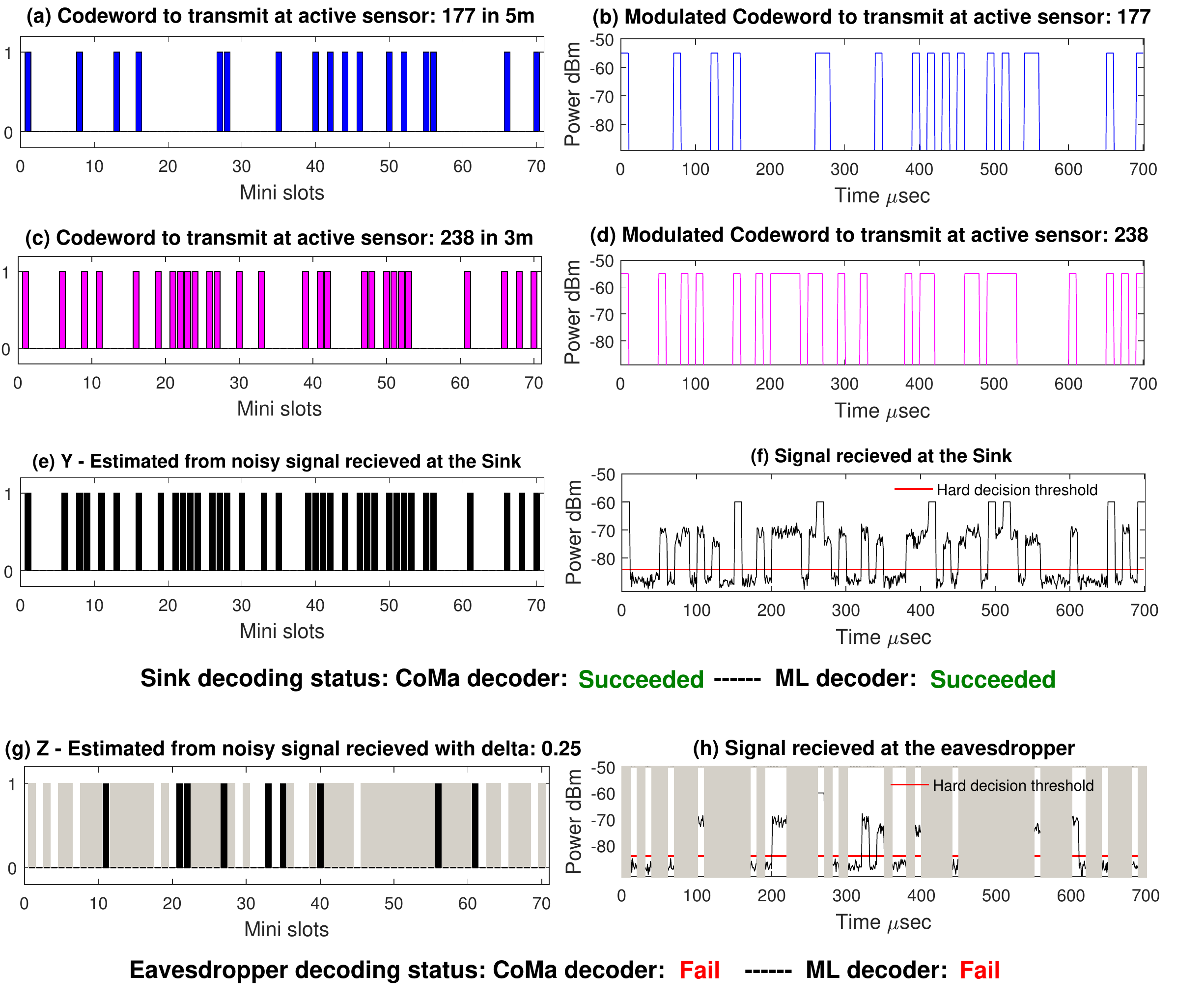}
  \caption{WSN Simulation in typical wireless environment.}
  \label{fig:Simulation_Real_Channel}
  \vspace{-1mm}
\end{figure}
\else
\begin{figure}
  \centering
  \includegraphics[trim= 6.9cm 1.95cm 19cm 1.6cm,clip,scale=0.370]{sim_ex_8-eps-converted-to.pdf}
  \caption{WSN Simulation in typical wireless environment.}
  \label{fig:Simulation_Real_Channel}
  \vspace{-1mm}
\end{figure}
\fi                 %
\fi
\end{document}